\documentclass[11pt]{article}

\usepackage[top=1in, bottom=1in, left=1in, right=1in]{geometry}

\usepackage{graphicx}
\usepackage{float}
\usepackage[space]{grffile}

\usepackage[tight]{subfigure}

\usepackage{amsmath}
\usepackage{amsthm}
\usepackage{amsfonts}
\usepackage{amssymb}

\usepackage{cite}
\usepackage[numbers]{natbib}

\newcommand{\comment}[1]{}

\title{Semi-supervised network inference with time course data and dynamic regression simulation}
\author{Phan Nguyen and Rosemary Braun}
\date{\today}

\begin{document}

\maketitle

\begin{abstract}
\noindent\textbf{Motivation: }Inferring the structure of gene regulatory networks from high--throughput datasets remains an important and unsolved problem.  Current methods are hampered by problems such as noise, low sample size, and incomplete characterizations of regulatory dynamics, leading to networks with missing and anomalous links.  Integration of prior network information (e.g., from pathway databases) has the potential to improve reconstructions.\\
\textbf{Results: }We developed a semi--supervised network reconstruction algorithm that enables the synthesis of information from partially known networks with time course gene expression data. We adapted PLS-VIP for time course data and used reference networks to simulate expression data from which null distributions of VIP scores are generated and used to estimate edge probabilities for input expression data. By using simulated dynamics to generate reference distributions, this approach incorporates previously known regulatory relationships and links the network to the dynamics to form a semi-supervised approach that discovers novel and anomalous connections. We applied this approach to data from a sleep deprivation study with KEGG pathways treated as prior networks, as well as to synthetic data from several DREAM challenges, and find that it is able to recover many of the true edges and identify errors in these networks, suggesting its ability to derive posterior networks that accurately reflect gene expression dynamics. \\
\textbf{Availability: }R code is available at https://github.com/pn51/postPLSR. \\
\textbf{Contact: }rbraun@northwestern.edu \\
\end{abstract}


\section{Introduction}

For cells to function properly, thousands of genes must interact in a concerted effort to produce the appropriate amounts of RNA and protein required for a variety of biological processes~\cite[]{PMC3083081,PMID18797474,PMID26355593,iyer2017computational} The collection of genes, their products, and interactions between them comprise gene regulatory networks (GRNs) which regulate their abundance and activity. Understanding the dynamics and structure of these networks can shed light on the regulatory cascades responsible for the emergence of different phenotypes, disease mechanisms, metabolic processes, and other biological functions.

One way to elucidate the interactions between genes is the use of microarray and sequencing assays that can measure the activity levels of thousands of genes simultaneously. Advances in high-throughput technologies have enabled the generation and widespread availability of rich datasets in an affordable and efficient manner. One of the major goals in functional genomics and systems biology is the prediction of functional relationships between genes from these datasets via computational means~\cite[]{PMID11911796,PMID24726980}. This procedure, gene network reconstruction, can offer new experimental directions to verify novel interactions, identify deficiencies in currently known networks and models, and understand how these networks function and can be perturbed to affect disease pathways and other vital processes.

While high-throughput sequencing techniques now generate large datasets efficiently and affordably, constructing accurate GRNs from these measurements remains a challenge. Typically, sample sizes are small compared to the number of genes measured. Consequently, network reconstruction is an underdetermined problem in which many models fit the data and an exponentially large space of networks needs to be considered. Furthermore, problems such as the stochastic nature of gene expression, experimental noise, missing data, difficulties in distinguishing between direct and indirect effects, and incomplete characterizations of the gene regulatory dynamics hinder the efficacy of many network reconstruction approaches. To produce accurate GRNs, algorithms need to address these issues with plausible, accurate modeling assumptions and constraints.

Many methods have been proposed to address these challenges. Early methods for GRN reconstruction used coexpression between gene expression profiles to identify relationships between genes and treat quantities such as correlation and mutual information as measures of edge confidence~\cite[]{PMID15486043,PMID10566452,PMID10902190}. CLR~\cite[]{PMID17214507}, and ARACNE~\cite[]{PMID16723010} built on mutual information-based relevance networks by filtering out indirect interactions. Regression-based methods with stability selection to control for false discoveries have also been adapted to estimate regulatory strength between genes~\cite[]{PMID23173819,vanSomeren2006}. Other approaches to determine causality between genes have been based on neural networks~\cite[]{PMID10380190}, probabilistic graphical models~\cite[]{PMID14764868}, Boolean networks~\cite[]{PMID5343519}, random forests~\cite[]{PMID20927193}, and partial least squares regression~\cite[]{PMID18204062,PMID26057728,PMID28031031}.

In general, these methods assume that data samples are independent in order to infer edges by using similarity- and causality-based edge confidence measures or by incorporating these samples into regression-type methods to estimate the influence among genes on their expression. Furthermore, early work in GRN reconstruction has focused on static data. However, since gene expression regulation is a dynamic process that can exhibit high autocorrelation, temporal data can be used to detect periodicity, identify cascades of differential expression, observe temporal responses to knockouts and other perturbations, and study expression evolution across different environmental, phenotypic, and other conditions, all of which may be used to infer causality between genes~\cite[]{PMID15130923}. Many GRN reconstruction methods have been developed to handle some of the features and challenges that are unique to time course data, some of which are adaptations of static methods, such as TD-ARACNE~\cite[]{PMID20338053}. Other methods include those based on Granger causality~\cite[]{PMID24067420}, dynamic time warping~\cite[]{PMC4816347}, dynamic Bayesian networks~\cite[]{PMID14534183}, and differential equations~\cite[]{PMID16686963,PMID19686586}.

Despite varying levels of success, many of these methods do not take advantage of known regulatory dependencies that are available in databases such as the Kyoto Encyclopedia of Genes and Genomes (KEGG)~\cite[]{PMID10592173,PMID24214961}. Several methods have been developed to leverage prior information in order to reduce false positive rates and improve network reconstructions. CoRe~\cite[]{PMID25910697} uses a known network and supervised learning to learn a classification model for each regulator of the network with the expression data of a potential target as input and edge confidence as output. The edge confidence scores are then recalibrated based on models learned after permuting the gene labels of the expression data and degree-preserving randomizations of the known network. iRafNet~\cite[]{PMID26072483} extends the random forest-based algorithm GENIE3~\cite[]{PMID20927193} by using prior regulatory information to bias the selection of known regulators when learning a random forest model for each target gene in a network.

Several methods to detect missing and anomalous links based on the network's topological features have also been proposed~\cite[]{Lu2011,PMC2799723,PMID18451861}. Typically, these methods compute edge likelihoods based on structural and generative assumptions or derive association quantities based on node degrees, common neighbors, path ensembles, and other network-based features~\cite[]{Lu2011}. However, they do not take into account any biological assumptions or experimental data. Given the availability of both pathway databases and expression datasets, it is therefore of interest to develop hybrid approaches that integrate transcriptomic data with partially known networks and accurate modeling constraints in order to refine these networks by detecting discrepancies and novel relationships.

In this paper, we describe a semi-supervised approach that integrates existing GRN model topologies with new time course gene expression data to infer novel interactions that are not captured in pathway databases. Our approach builds on PLS-VIP~\cite[]{Chong2005}, a variable selection method using partial least squares regression. Here, we adapt PLS-VIP for time course data by assuming that the expression of a gene is a function of the expression of the genes at the previous time point and calculate the VIP scores for the lagged PLSR model. Furthermore, we use a reference network derived from a pathway database to simulate expression data from which additional VIP scores are computed to generate a pair of null distributions for each ordered gene pair. The data-derived VIP scores are then compared to these respective null distributions to estimate posterior edge probabilities. In Section~\ref{sec:materials}, we describe this approach in more detail. In Section~\ref{sec:results}, we apply the method to time course datasets from past DREAM challenges, as well as to an experimental dataset from an insufficient sleep study using KEGG pathways as prior networks.  
The synthetic DREAM challenge data provides a highly controlled framework in which to evaluate the performance of our approach, whereas application to the experimental dataset and KEGG pathways enables us to compare our method against competing approaches in real data.
While the connections specified in KEGG pathways have been expertly curated, we expect that there remain innaccuracies (either missing edges that have yet to be experimentally identified, or anomalous links due to prior false positive results) that our method is designed to detect.
In both applications, we show that the method is able to recover significant portions of these networks while also being able to identify novel and anomalous connections. These results suggest the method's ability to derive posterior networks that accurately reflect gene expression dynamics by incorporating previously known regulatory relationships and linking the network to the dynamics. In Section~\ref{sec:conclusion}, we conclude and discuss possible extensions to our approach.


\section{Methods}\label{sec:materials}

\subsection{Background: PLSR-based methods}

In general, regression-based methods assume that the expression of a gene can be modeled as a function of all other genes in a dataset,
and determine if an edge exists using the magnitude of the association a predictor and the target gene.
Multiple linear regression can be used to model the relationship between a set of predictors and a response when the number of predictors is
low relative to the number of available samples and when the predictors are not highly collinear. However, gene expression and
other biological datasets are subject to problems with low sample sizes, high dimensionality, and multicollinearity, making multiple linear regression inappropriate for modeling relationships in those
datasets. One approach that has been used for predictive modeling while also accounting for these problems is partial least squares
regression (PLSR)~\cite[]{Wold2001,Hoskuldsson1988,Chong2005}. PLSR
seeks to simultaneously decompose predictor and response matrices with sets of latent variables for each matrix such that the covariance
between the sets of latent variables is maximized.  This procedure permits the high dimensionality of the gene expression space to
be reduced to a smaller number of PLSR components, removing latent sources of variance that have little predictive value.

Several GRN reconstruction approaches have used PLSR as an underlying method. \cite{PMID18204062} introduced an unsupervised method that assumes that the expression of each gene is a linear function of the expression of the remaining genes in a dataset. PLSR is then used to construct a model for each gene, and an interaction score between two genes is derived based on the contribution of the latent variables to the predictor gene and on the contribution of the predictor gene to the latent variables of the target gene's model. These scores are then aggregated and thresholded to predict an undirected network. The method was shown to be able to identify potential interactions, but it exhibited high false positive rates, required large sample sizes for better quality results, and only predicted undirected edges that corresponded to association rather than causation.

The PLSR ``variable importance in projection'' (VIP) score provides an efficient means to select predictor genes that most strongly predict the target gene, between which edges may be inferred to exist.
Briefly, the VIP score quantifies the contribution of each gene to the latent variables as the weighted sum of the squared correlations between the PLSR components and the original variable~\cite[]{Chong2005}.
A variable with a large VIP score (close to or greater than 1) can be considered important in given model, and thus
PLS-VIP may be used to identify predictor genes that most strongly influence the target gene.
A number of methods have used PLS-VIP to infer edges in GRNs.
DIONESUS~\cite[]{PMID26057728} is an unsupervised GRN reconstruction algorithm that models the expression of each gene as a function of the expression of the remaining genes.  It iterates through the genes of a dataset, treating each as a response and the remaining genes as potential predictors, and uses PLS-VIP to compute VIP scores for the predictors. These scores are treated as measures of edge confidence, with higher scores assumed to be more indicative of an edge, and are aggregated and thresholded to predict directed edges. DIONESUS was shown to be scalable, efficient, and accurate on \textit{in silico} data and was applied to microwestern array and cell viability assay data to reconstruct a cell signaling network in a human carcinoma cell line driven by the overexpression of Epidermal Growth Factor Receptor, and was also successful in reconstructing networks in several DREAM challenges. PLSNET~\cite[]{PMID28031031} is a variant of the PLS-VIP approach that incorporates sample-bagging and feature-bagging. At each iteration, the algorithm applies PLS-VIP to a bootstrapped expression dataset, and the VIP scores for each regulator-target pair are added across all iterations and rescaled to account for a regulator's influence across different target genes.

Our method extends these ideas in two novel and important ways.  First, by using time--course gene expression data, we are able to make causal inferences about gene regulation, permitting the reconstruction of directed networks that reflect regulatory dependencies.  Second, by simulating gene expression dynamics from known networks, our method incorporates previously known regulatory relationships in a semi-supervised manner and links the network structure to the dynamical behavior of the system.


	\subsection{Extending PLSR to time--course data}

Since gene expression regulation is a dynamic process, time course data can be measured and used to infer causality, identify activated genes and changes in differential expression, detect periodicity, determine coexpression, and provide insight into other temporal aspects and mechanisms that cannot be ascertained from static data. One basic approach that has been used to model and analyze time course expression data, such as in methods based on Granger causality and Markov processes, is to assume that the data can be modeled with vector autoregression~\cite[]{PMID24067420,PMID11793246}. More specifically, the one-way temporal ordering of the data can be exploited by assuming that the expression of a gene is linearly dependent on the expression of its regulators at a previous time point. A lag of one time interval between the predictor and response variables is typically used, so that the expression of a gene is described by
	\begin{equation} \label{eq:exprModel}
		x_i\left(t + \Delta t\right) =  \sum_{j \neq i} w_{ji} x_j\left(t\right) + \epsilon\,,
	\end{equation}
where $x_i\left(t\right)$ is the expression of gene $i$ at time $t$, and $w_{ji}$ are weights indicating how much the expression level of gene $j$ influences that of gene $i$ over a time interval $\Delta t$.

Gene expression has more typically been modeled with differential equations~\cite[]{PMID15169868,PMID10659856,PMID12843395, PMID20308593}, particularly when the interest is in how gene expression changes during a biological process or as a result of perturbations and stimuli. In the linear case,
	\begin{equation*}
		\frac{d x_i\left(t\right)}{dt} =  \sum_{j \neq i} a_{ji} x_j\left(t\right) + \epsilon\,,
	\end{equation*}
where $a_{ji}$ is the rate of influence of the expression of gene $j$ on that of gene $i$. Since microarray data can only collected at discrete time points, discretizing results in
	\begin{equation} \label{eq:deltaExprModel}
		\Delta x_i\left(t+\Delta t\right) =  \sum_{j \neq i} w'_{ji} x_j\left(t\right) + \epsilon\,,
	\end{equation}
where $w'_{ji}= a_{ji}\Delta t$ and $\Delta x_i\left(t+\Delta t\right) = x_i\left(t+\Delta t\right) - x_i\left(t\right)$.
PLSR may be applied to fit (\ref{eq:exprModel}) or (\ref{eq:deltaExprModel}) to time--course gene expression data.

To assess the contribution of a predictor gene $j$ on a target gene $i$, we compute the VIP score $v_{ji}$ for $j$ in the expression model for $i$. However, rather than setting a common threshold $v_\text{thresh}$ and assigning edges when $v_{ji}>v_\text{thresh}$, we instead compare $v_{ji}$ to the distributions of VIP scores that would be expected if the edge is/isn't present using simulated network dynamics.  This approach has the appealing feature of integrating known regulatory links into the analysis and provides a means to assign a posterior probability to a given edge, as detailed below.

	\subsection{Semi-supervised network reconstruction}

Partial knowledge of GRNs is available in pathway databases such as KEGG. Given the availability of these knowledge-derived networks and transcriptomic datasets, it is useful to consider integrating this information in GRN reconstruction, using experimental data to identify novel or anomalous edges from existing networks.  In contrast to many existing GRN reconstruction methods that construct \textit{de novo} networks solely from expression data, semi-supervised methods can incorporate a partially known network with accurate modeling constraints and dynamics in order to explain the observed gene expression values and identify deficiencies that contribute to discrepancies between the observed expression values and the dynamics of the partially known network.

Here, we use prior networks from pathway databases to inform the inclusion of edges in the inferred networks.
To motivate our approach, it is instructive to consider some of the properties of PLS-VIP and prior PLS-VIP-based approaches. In DIONESUS, a \textit{de novo} network is constructed based on the intuition that because a VIP score is a measure of a predictor's contribution to a PLSR model, a high VIP score for a predictor-response pair should be treated as evidence for the existence of a corresponding edge in the network. Using this assumption, the method aggregates the VIP scores across the PLSR models for all genes and imposes a cutoff to determine the edges of the network. However, the mathematical properties of the VIP scores---namely, that the mean of the squares of the VIP scores within each gene's model is equal to 1---can contribute to a large number of errors when multiple models are aggregated and thresholded.
If a gene is affected by most or all of the genes in a pathway, the VIP scores will summarize the relative importance of these regulators to each other, but when combining the scores and imposing a standard VIP cutoff of 1, many of those edges will not be identified. Similarly, when a gene has few regulators, the same procedure will identify many false positive edges. Therefore, while VIP scores from the same model may be compared to each other to determine the relative importance of potential predictors of the same target gene, they are not necessarily comparable across regression problems corresponding to different target genes.

Rather than comparing VIP scores across PLSR models for different genes using a common threshold, we compare the VIP score for a pair of genes to distributions of scores that would be expected of that pair given the structure of the rest of the network and a dynamic model for gene expression. To obtain these distributions, we use the known regulatory connections of a pathway along with the dynamic model to simulate time course data from which reference distributions of VIP scores are computed. By comparing the data-derived VIP scores to those derived from simulated data that is potentially observable based on the network and the assumed dynamics on that network, we can identify VIP scores that are atypical for a pair of genes and may correspond to novel or anomalous edges.

\begin{figure}
	\centering
	\includegraphics[width=.97\textwidth]{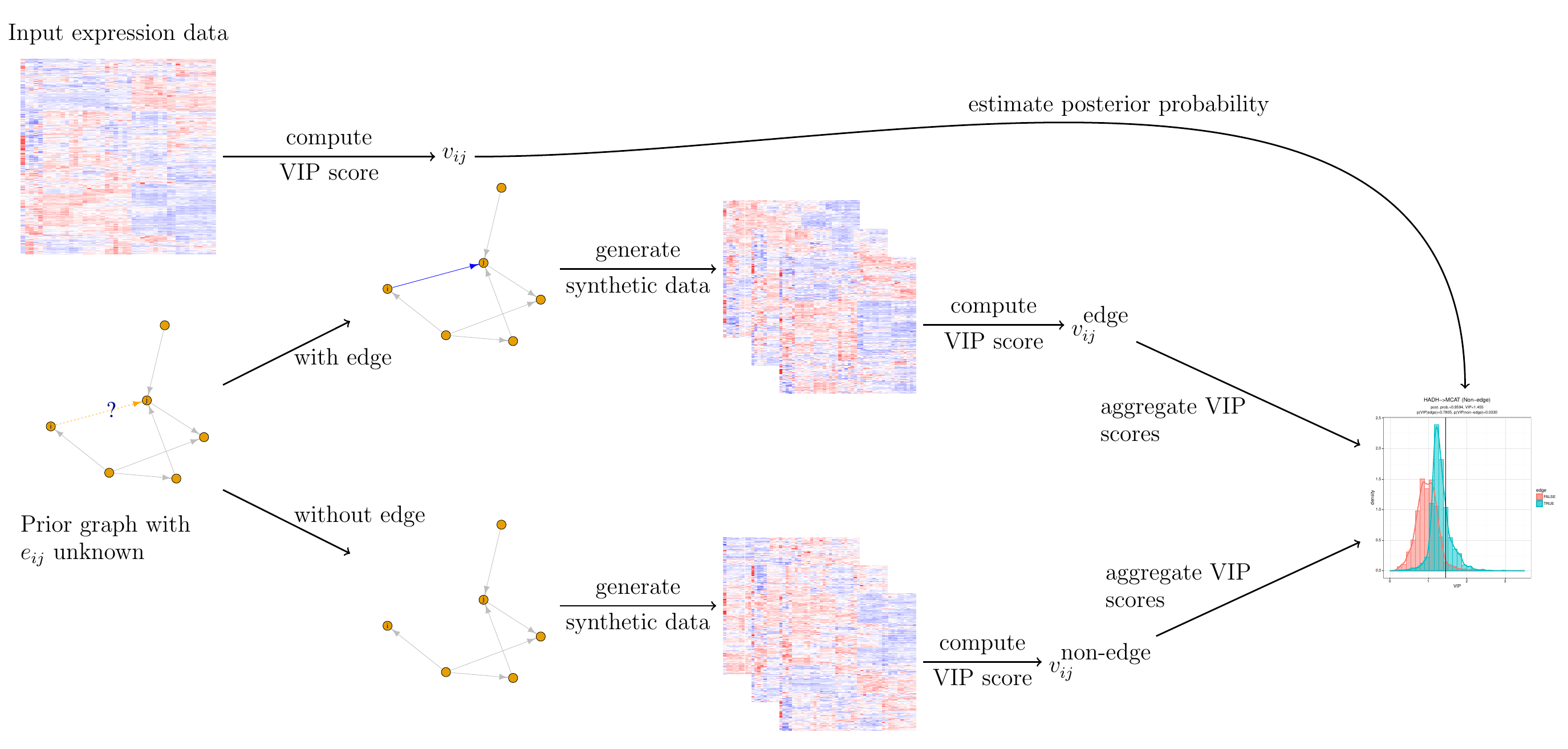}
	\caption{Workflow for the semi-supervised gene network reconstruction approach. The inputs are an observed time-course expression dataset and a network obtained from a pathway database.  For the pair $(ij)$ of interest, synthetic expression data are obtained from simulated dynamics on the network with and without edge $e_{ij}$. The VIP score for $(ij)$ is computed from the input expression data and compared to the distributions of VIP scores that would be expected with and without $e_{ij}$ from the simulations to obtain the posterior probability that $e_{ij}$ is an edge.\label{fig:methodWorkflow}}
\end{figure}

\begin{figure}
	\centering
	\includegraphics[width=.97\textwidth]{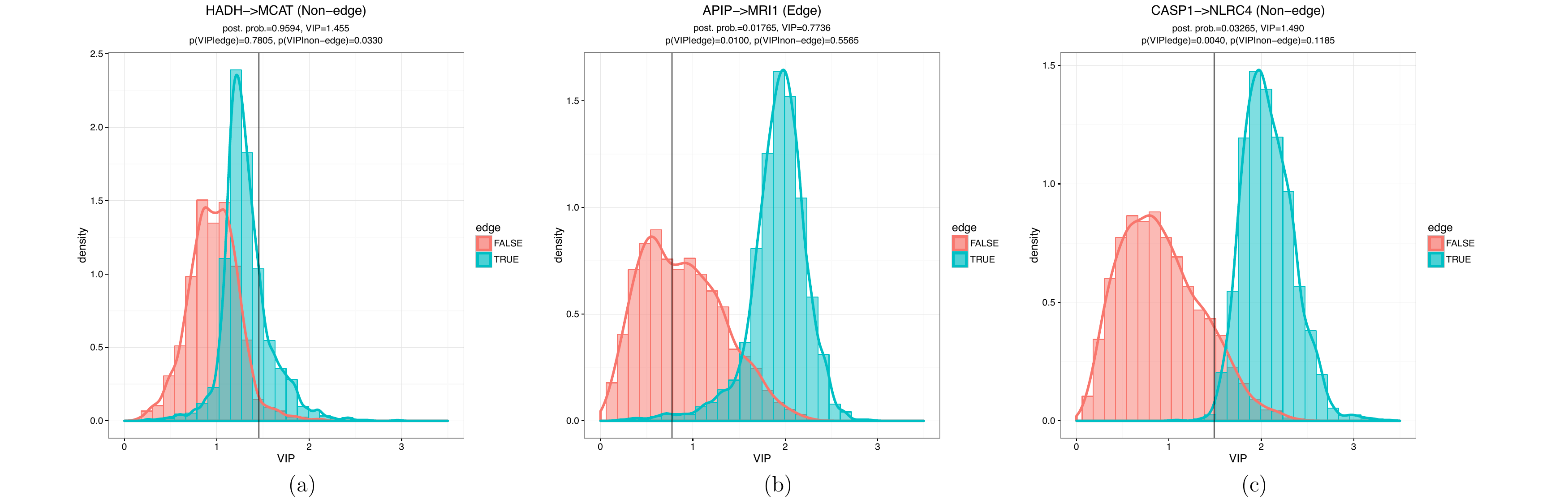}
	\caption{Examples of pairs of distributions generated by the semi-supervised PLS-VIP-based approach. (a) \textbf{Potential novel edge.} The data-derived VIP score for the prior non-edge is high relative to the pairs of reference distributions of VIP scores, which contributes to a high posterior edge probability. The non-edge represents a potential novel edge in the network. (b) \textbf{Potential anomalous edge.} Since the data-derived VIP score for the prior edge is low relative to the pairs of reference distributions of VIP scores, the posterior edge probability will be low. This represents a potential anomalous edge. (c) \textbf{Prior non-edge with a high VIP score but low posterior edge probability.} Using a VIP threshold would identify a false positive edge with a high VIP score, whereas the semi-supervised approach assigns a low posterior edge probability because the VIP score is low relative to its reference distribution of edge VIP scores. \label{fig:distributionExamples}}
\end{figure}

A summary of the overall approach is shown in Figure~\ref{fig:methodWorkflow}.
First, we assume that the expression of a gene can be modeled using (\ref{eq:exprModel}) or (\ref{eq:deltaExprModel}) and calculate the data-derived VIP scores for the lagged PLSR model and an input expression dataset. We then use the same model and the reference network to simulate expression data from which additional VIP scores are computed to generate a pair of reference distributions for each ordered gene pair: one assuming the reference network includes the edge of interest, and another with the same edge excluded from the network. For each of the two networks, 2000 gene expression trajectories were simulated using randomly assigned weights $w_{ji}$ for every edge, and the VIP scores were computed from these simulated trajectories.
Figure \ref{fig:distributionExamples} shows examples of these simulated pairs of VIP distributions.
The posterior edge probabilities are then estimated by comparing the data-derived VIP scores (shown as black vertical lines in Figure \ref{fig:distributionExamples}) to these respective distributions, and a network can be determined by thresholding the probabilities. We identify novel connections by prior non-edges with high posterior probabilities and erroneous edges by prior edges with low posterior probabilities. By using simulated dynamics to generate reference distributions, this approach incorporates previously known regulatory relationships and links the network to the dynamics to form a semi-supervised approach that uses the prior network and expression data to recover the true edges of the known network and discover novel and anomalous connections.

In our applications, we assume that gene expression is described by (\ref{eq:exprModel}). Additional mathematical details, details about the posterior edge probability estimation, and analysis of the parameters are available in the Supplementary Information.

	\subsection{Datasets}

		\subsubsection{DREAM}

We applied our method to synthetic time course gene expression data from several DREAM challenges. In one of the DREAM2 challenges, 50-node networks were derived from Erdos-Renyi and scale-free topologies with Hill-type kinetics driving gene expression~\cite[]{PMID19348640}. The DREAM3 \textit{in silico} network challenge contained 10-, 50-, and 100-gene subnetworks extracted from \textit{E. coli} and \textit{S. cerevisiae} gene network with expression values simulated using GeneNetWeaver~\cite[]{PMID20186320, PMID19183003, PMID20308593}. Finally, in the DREAM4 \textit{in silico} network challenge, GeneNetWeaver was used to apply various perturbations to 10- and 100-gene networks and the network response was measured before and after the perturbations were removed.

		\subsubsection{Insufficient sleep}

We also applied our method to a time course microarray dataset from a study of the mechanisms and effects of insufficient sleep and circadian rhythm disruption on gene expression, circadian regulation, and other related processes~\cite[]{PMID23440187}.
Data was collected by subjecting 26 participants to restricted sleep and control conditions, each followed by an extended period of constant routine during which blood samples were periodically collected for RNA extraction. Since many time points and samples are available, the richness of the dataset makes it amenable to many types of analyses.

To apply our method to the insufficient sleep dataset, we treat KEGG pathways as reference networks. In particular, we only consider subgraphs consisting of genes for which expression data is available. Since there are edges that appear in one pathway and not another between the same pair of genes, we first merge all of the pathways together. With the resulting graph and for each pathway, we take the induced subgraph consisting of genes that are in both the pathway and expression dataset. Finally, since this procedure may leave many singleton nodes in the subgraph, we then take the largest component of the subgraph and use it as the input network.


\section{Results}\label{sec:results}

	\subsection{DREAM}\label{subsec:dreamApp}

	\subsubsection{Network/edge recovery with synthetic data}

	We first evaluate the method by applying it directly to the DREAM datasets without any modifications to the networks. An area under the curve (AUC) can be computed by sorting the posterior edge probabilities and calculating the true-- and false--positive rates as a function of the posterior probability threshold for including an edge. When using the original network as the input and reference for the AUC computation, the AUC can be interpreted as a network recovery rate, which will be close to 1 if the method is accurate and there are very few novel or anomalous edges in the network. Since the DREAM datasets are synthetic and the networks are completely known, the method should ideally assign high and low edge probabilities to the edges and non-edges, respectively, in order to return AUC values that are close to 1. If the input network is known to be incomplete, high AUC values are still desirable as an indicator that the regulatory connections that we know about are indeed correct. Furthermore, when a majority of the edges (non-edges) are assigned high (low) posterior probabilities, the high (low) edge probabilities that are assigned to non-edges (edges) in the input network can be treated as putative evidence for a corresponding novel (anomalous) edge in the network, since the probabilities will rank similarly with those of the true edges (non-edges).

\begin{figure}
			\centering
			\includegraphics[width=\columnwidth]{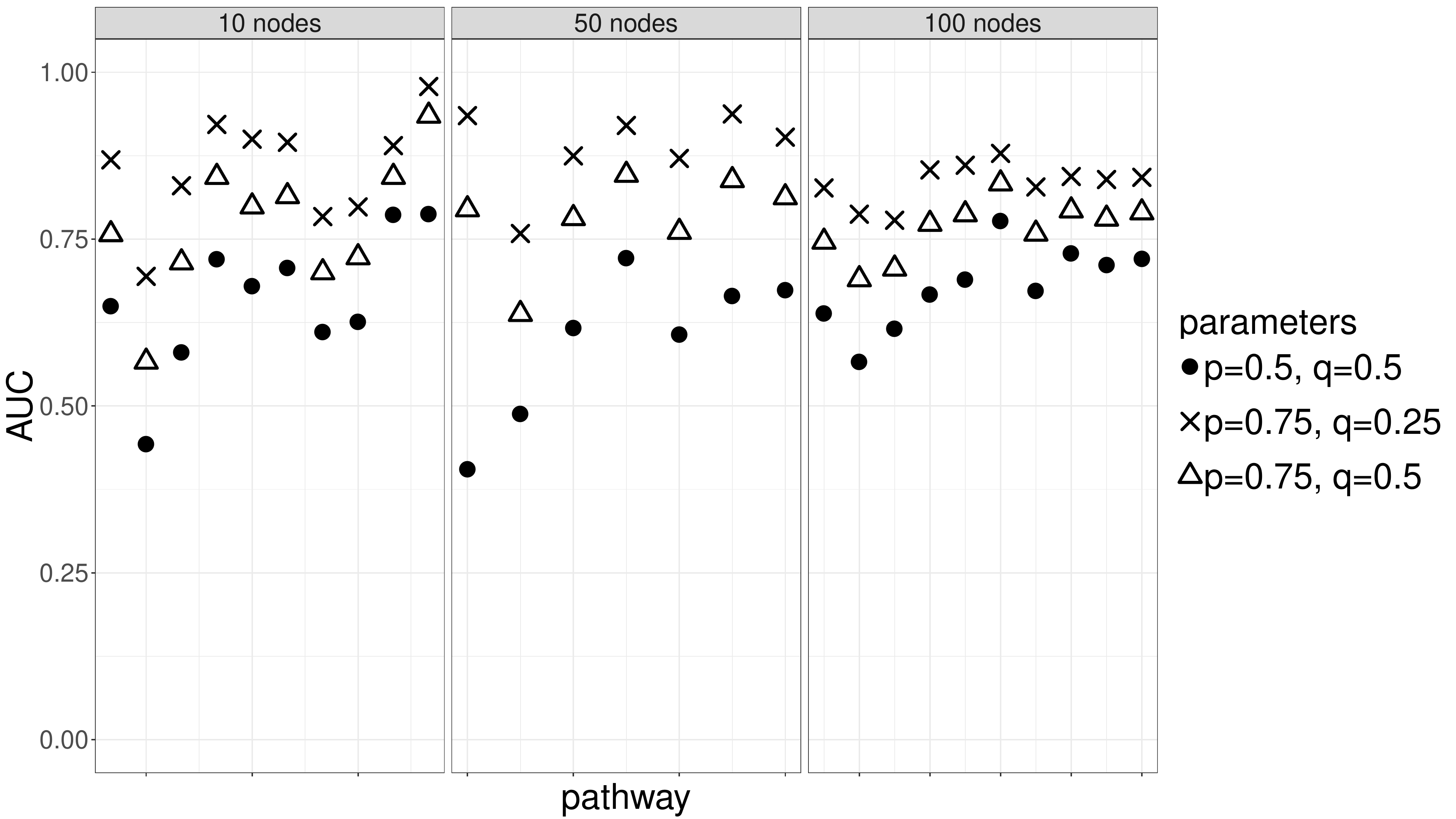}
			\caption{AUCs for the posterior PLS-VIP-based method when applied to the DREAM datasets.
			Pathways are arranged along the $x$-axis by their number of nodes; 10 networks contain 10 nodes, 7 networks contain 50 nodes, and 10 networks contain 100 nodes.
			\label{fig:dreamAUC}}
		\end{figure}

	In Figure~\ref{fig:dreamAUC}, the DREAM dataset AUCs are shown for prior probabilities $(p,q) \in \lbrace (.5,.5),\allowbreak\, (.75,.25),\, (.75,.5) \rbrace$. Even when uninformative priors $p=q=.5$ are used, except in a few cases, the method returns relatively high AUC values. When $p$ increases or $q$ decreases from those values, the AUCs also increase; as $p$ increases, the posterior edge probabilities of the prior edges will increase, and when $q$ decreases, the posterior edge probabilities of the prior non-edges will decrease, both of which contribute to a higher AUC. While this suggests using large values of $p$ and small values of $q$, some care must be taken in choosing these parameters, especially when one of the goals of reconstructing GRNs is discovering novel regulatory relationships between genes. For a potential novel edge to be proposed, a prior non-edge should have a posterior edge probability that is higher than those of the true non-edges and comparable to those of the edges in the graph. If $q$ is set close to zero or extremely low relative to $p$, then the posterior edge probabilities of all prior non-edges will be small compared to the those of the prior edges. In this case, the AUC will be 1, but no potential novel edges will be proposed.

	\subsubsection{Detection of novel and anomalous edges}

	While the previous analysis explains how posterior probabilities are generally assigned to edges and non-edges as a function of the prior parameters, how this affects the recovery of the input network, and why non-edges (edges) with high (low) posterior edge probabilities are possible novel (anomalous) connections, it says little about the method's ability to detect novel edges in a network. To evaluate its link prediction ability, we can remove some of the edges from the DREAM networks, run the method with the modified networks, and consider the highest ranking non-edges. For a numerical measure of this ability, an AUC can be calculated by ranking the posterior edge probabilities of the non-edges and removed edges. In this case, the AUC can be interpreted as the probability that a randomly chosen missing edge has a higher posterior edge probability than that of randomly chosen a non-edge. The degree to which the AUC exceeds 0.5 is a measure of how much better the algorithm does than chance, so values closer to 1 are indicative of better novel edge detection performance. (A graphical depiction of this proceedure is given in Section 2, Figure~A-1 of the Supplementary Information.)

	In Figure~\ref{fig:dreamRemovedAUC}, we randomly removed 4, 8, 16, and 32 edges over 20 iterations for each of the 50-node DREAM networks. Except for the first two networks, the method generally does better than chance at identifying the missing edges of the networks. In addition, even as the number of removed edges increases, the mean of the AUCs appears to be fairly stable. These observations suggest the method's potential to detect novel edges, and when many regulatory links are missing, the method is still able to use the expression data along with the dynamics on the rest of the partially known network to identify many of the missing edges that contribute to inconsistencies between the expression data and the prior network.

		\begin{figure}
			\centering
			\includegraphics[width=\columnwidth]{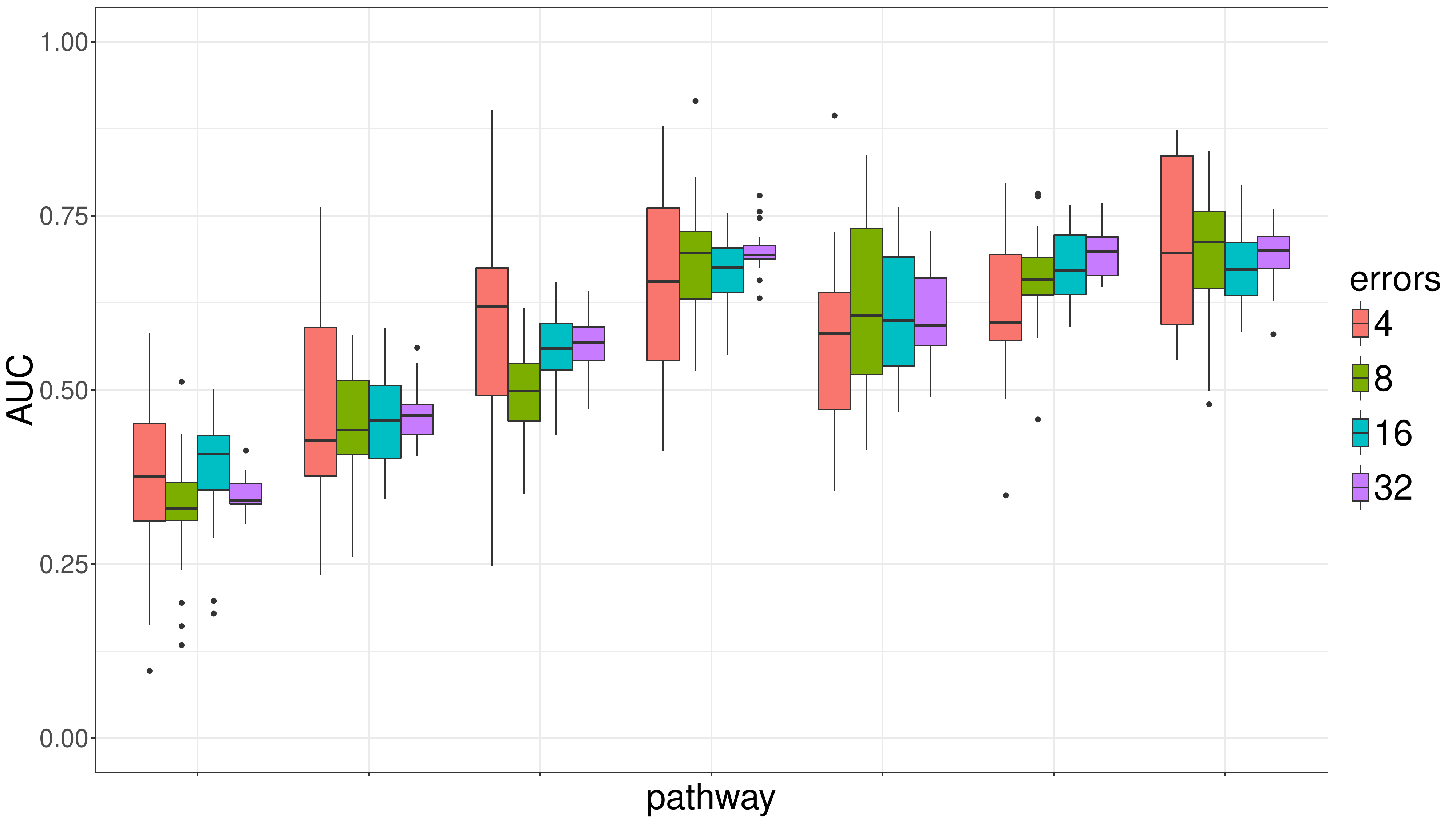}
			\caption{AUCs for the posterior PLS-VIP-based method when used for novel edge detection. 4, 8, 16, and 32 edges are randomly removed from the 50-node DREAM networks for 20 iterations, and AUCs are calculated using the non-edges of the modified networks. 
			\label{fig:dreamRemovedAUC}}
		\end{figure}

\begin{figure}
			\centering
			\includegraphics[width=\columnwidth]{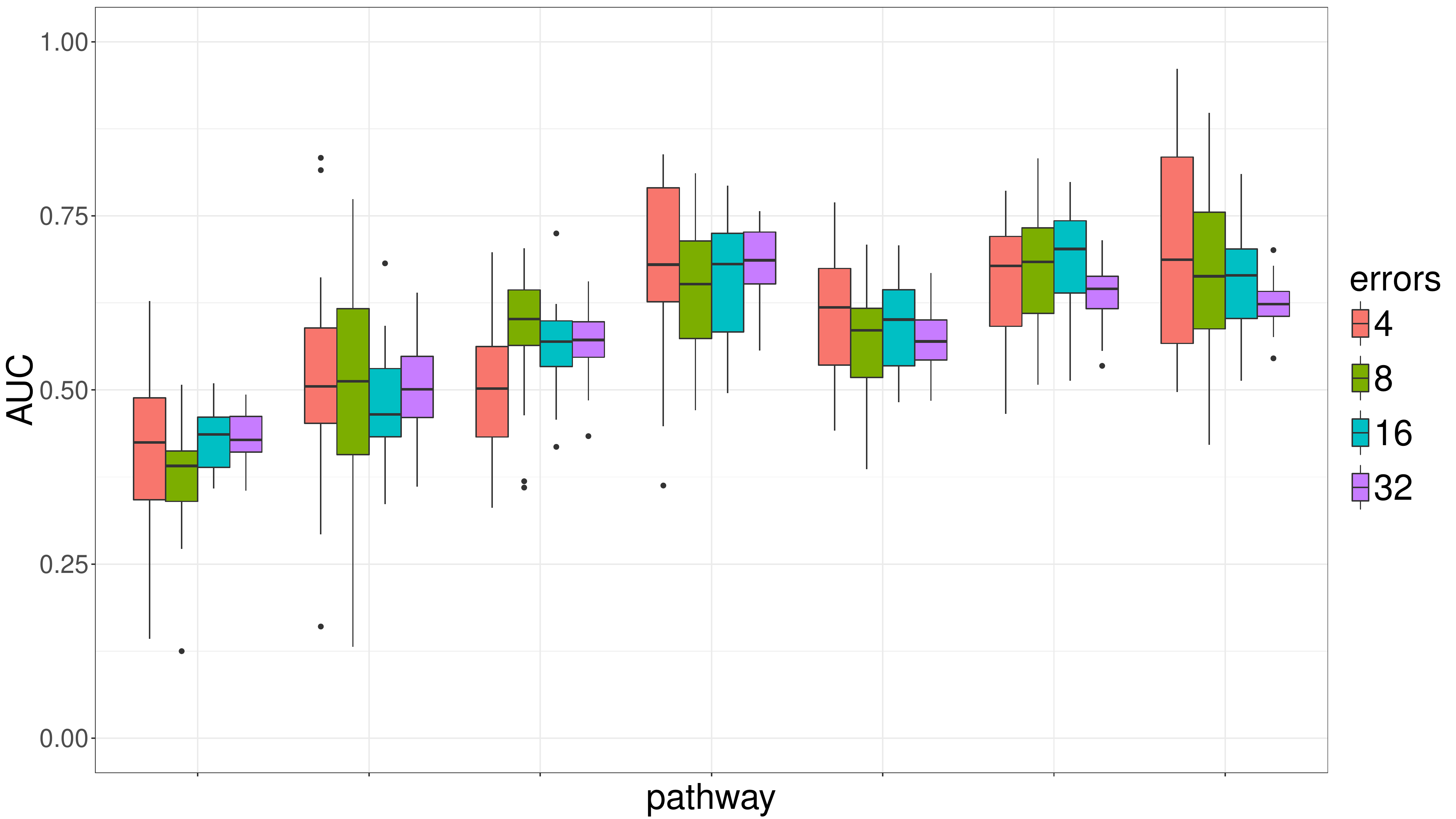}
			\caption{AUCS for the posterior PLS-VIP-based method when used for anomaly detection. 4, 8, 16, and 32 edges are randomly added to the 50-node DREAM networks for 20 iterations, and AUCs are calculated using the edges of the modified networks.
			\label{fig:dreamAddedAUC}}
		\end{figure}

	We can perform a similar analysis to evaluate the method's ability to detect anomalous edges. To do so, we can add edges to the networks, run the method with the modified networks, and look at the lowest ranking edges. We can calculate an AUC by ranking the posterior probabilities of the edges and added edges to summarize the anomaly detection performance. In this case, the AUC can be interpreted as the probability that a randomly chosen edge has a higher posterior edge probability than that of a randomly chosen false edge, and higher AUCs correspond to better performance. In Figure~\ref{fig:dreamAddedAUC}, we randomly added 4, 8, 16, and 32 edges over 20 iterations for each of the 50-node DREAM networks. As with the novel edge detection case, the method does not perform well on the first two networks, but the rest of the networks exhibit ranges of AUCs that are above 0.5. Based on this performance, the method can potentially be used for anomaly detection.

	\subsection{Insufficient sleep}\label{subsec:sleepApp}

	\subsubsection{Network/edge recovery}

	We now consider an application to the insufficient sleep dataset with KEGG pathways as input networks. Since these networks are known to be incomplete, the primary interest should be discovering novel edges, represented by prior non-edges with high posterior edge probabilities. In addition, while the interactions that comprise these pathways have been expertly curated, it is possible for some studies to contain false positive results, while other study results may be irreproducible. These correspond to anomalies, represented by prior edges with low posterior probabilities.

\begin{figure}
			\centering
			\includegraphics[width=\columnwidth]{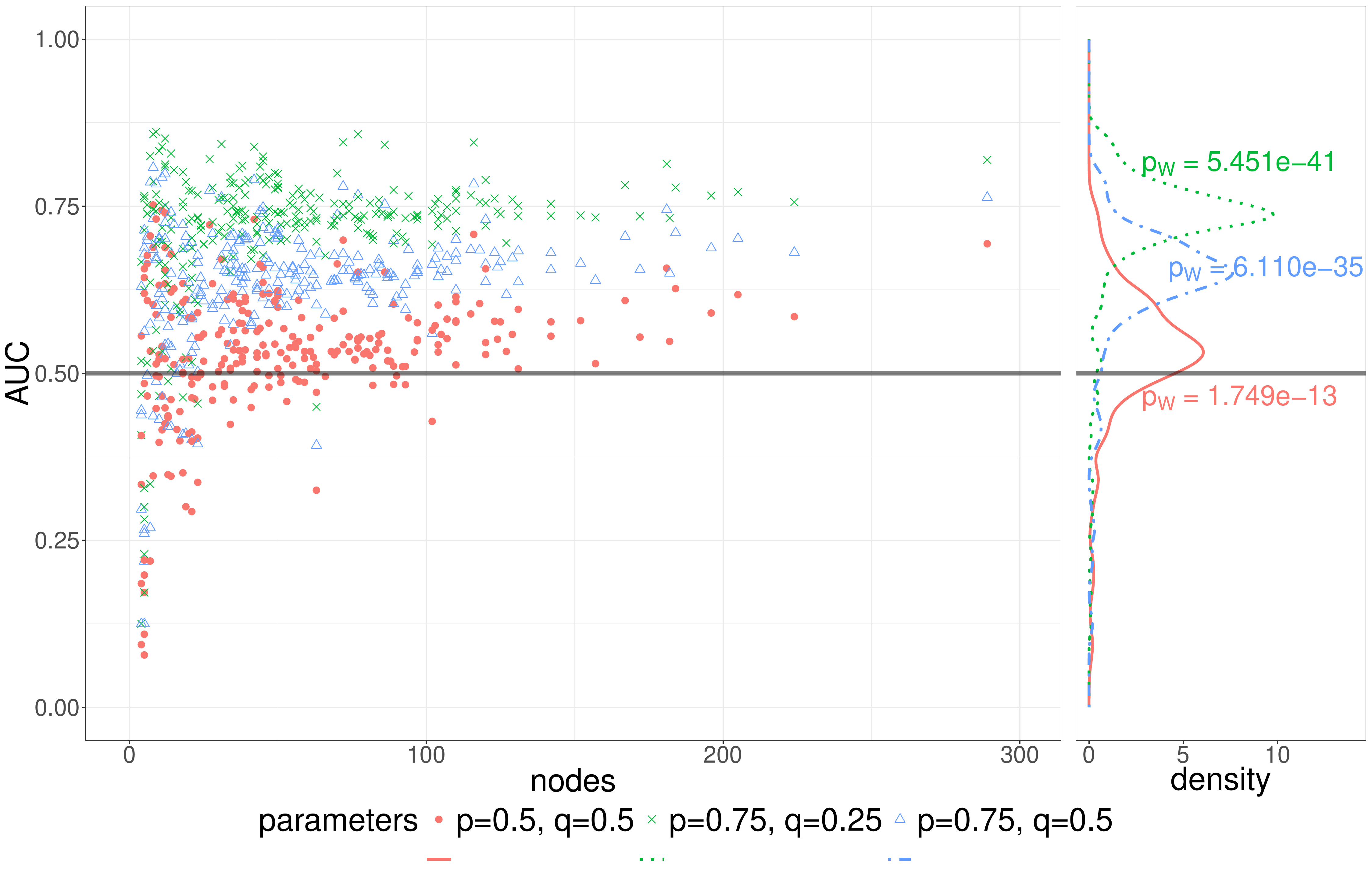}
			\caption{AUCs for the posterior PLS-VIP-based method when applied to the insufficient sleep dataset with KEGG pathways as known prior graphs. Wilcoxon test $p$-values for $H_0: \mu_{\text{AUC}} \leq 0.5$ and $H_1: \mu_{\text{AUC}} > 0.5$ are also shown for each set of prior parameters.
			\label{fig:sleepAUC}}
		\end{figure}

	As with the DREAM networks, we compute the AUCs with the KEGG pathways as reference networks using the same prior probability parameters in Figure~\ref{fig:sleepAUC}. Compared to the DREAM network AUCs at the same prior parameters, the AUCs in this case are lower, which may be a result of the networks being incomplete and the expression data being real. With $p=q=.5$, the method only appears to perform slightly but significantly better than chance at recovering the pathway edges, so it is likely to suggest many potential novel edges, most of which may be false discoveries. To reduce the number of false discoveries, a combination of higher $p$ and lower $q$ should be used. The former results in higher posterior edge probabilities being assigned to many of the prior edges, which will reflect higher confidence in the studies used to form the pathways but yield little to no candidate edges as anomalies. The latter will assign lower posterior edge probabilities to many of the prior non-edges, leaving very few non-edges with high posterior edge probabilties that can be proposed as novel edges and therefore reducing the number of false discoveries.

	\subsubsection{Posterior edge probability and VIP score comparison}

\begin{figure}
		\centering
		\includegraphics[width=\columnwidth]{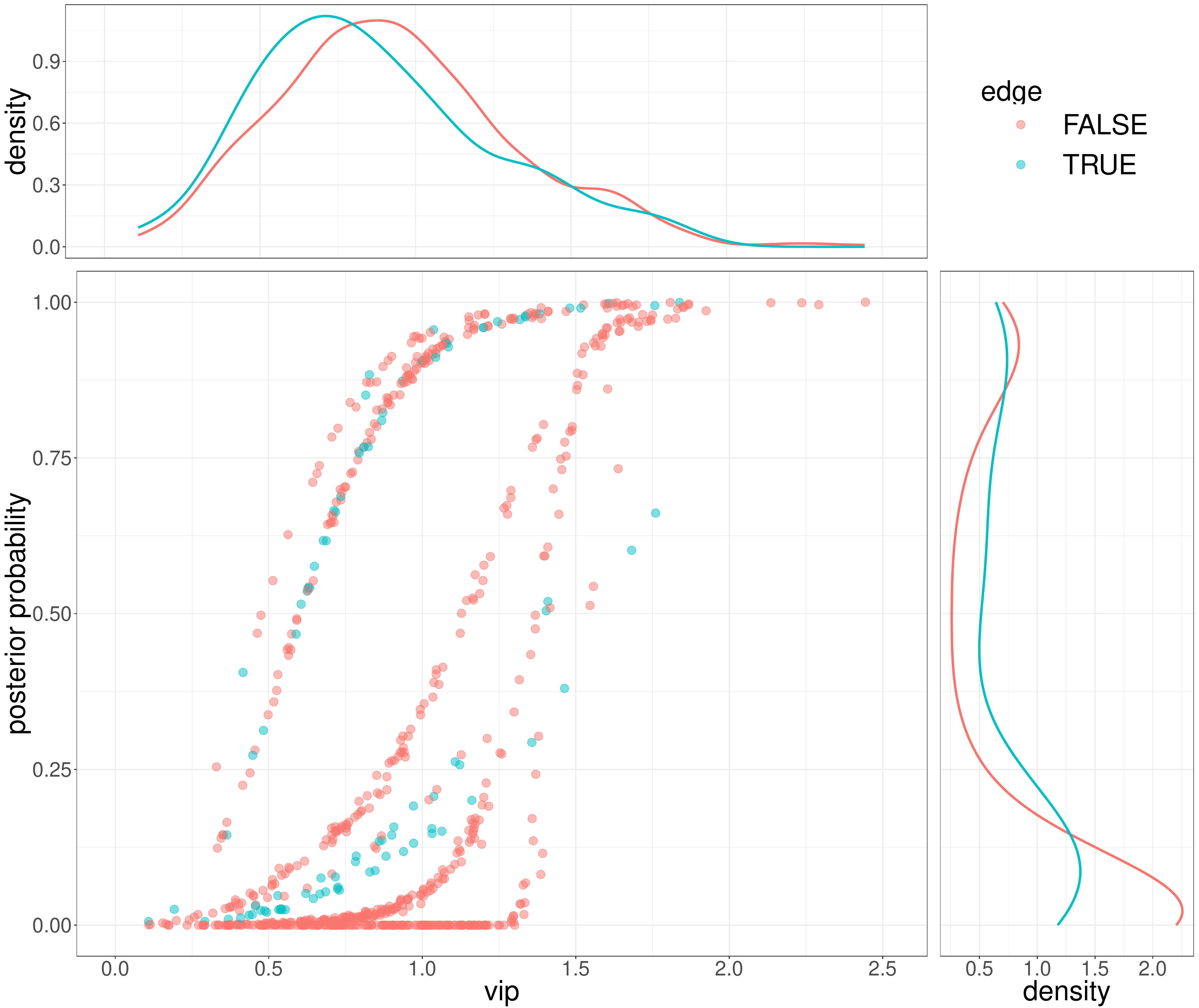}
		\caption{Posterior edge probability vs. VIP score for each pair of genes in the circadian rhythm pathway with $p=q=.5$. Large VIP scores tend to have higher posterior probabilities, but a small VIP can still result in a high posterior probability and vice-versa. 
		\label{fig:postVvip-marginal}}
	\end{figure}

	Since our method transforms VIP scores into posterior edge probabilities, it is useful to see how the computed posterior probabilities compare with the data-derived VIP scores. In Figure~\ref{fig:postVvip-marginal}, the posterior probabilities are plotted against the VIP scores for each ordered pair of genes in the circadian rhythm pathway using $p=q=.5$, colored by prior edge existence. We note that high (low) VIP scores do not necessarily correspond to high (low) edge probabilities. More specifically, many of the prior edges (cyan) have moderate to high edge probabilities, many having VIP scores that are below 1. Similarly, there are many non-edges (red) with relatively low edge probabilities, some having VIP scores that are greater than 1. Therefore, many of the edges and non-edges that would have been misclassified by directly comparing VIP scores are more likely to be classified correctly using the posterior probabilities. We also see that the distribution of VIP scores for the prior edges are slightly more skewed towards lower values than that of the prior non-edges. However, the corresponding posterior probabilities for the prior edges tend to cover moderate to high values, whereas those of the non-edges are bimodally concentrated around 0 and less so around 1. It can also be observed that the posterior edge probabilities for true edges are generally higher than those for non-edges at the same VIP scores, and as with the AUCs, further improvements can be made by adjusting the prior probabilities appropriately.  (The multiple curves visible in Figure~\ref{fig:postVvip-marginal} can be attributed to automorphically equivalent nodes; a detailed explanation may be found in the Supplementary Information.)

	\subsubsection{Method comparisons}

\begin{figure}
			\centering
			\includegraphics[width=\columnwidth]{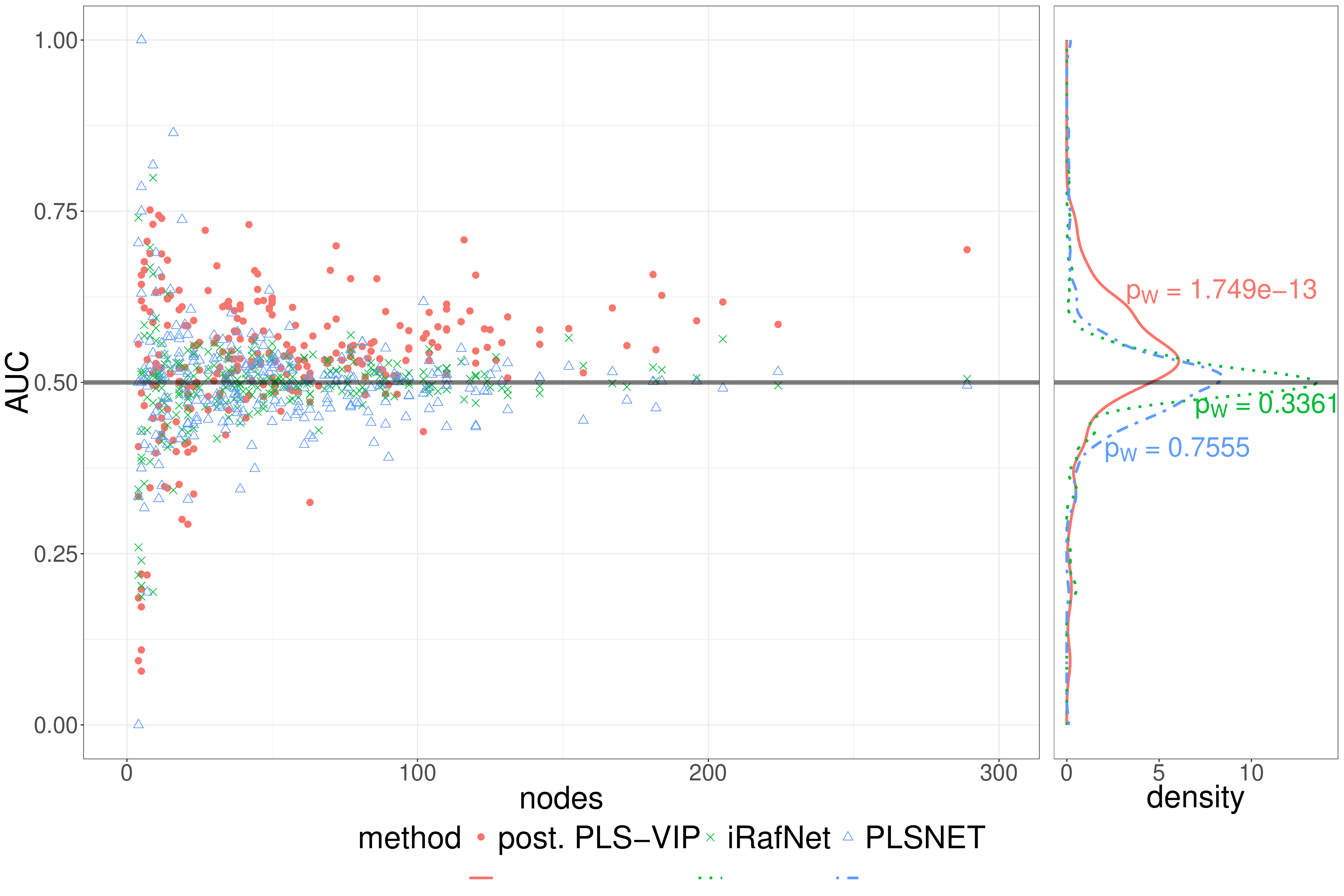}
			\caption{AUCs obtained by the proposed method (post PLS-VIP), iRafNet, and PLSNET for KEGG pathways using the sleep dataset.  Methods were applied using default and/or uninformative prior parameters. Wilcoxon test $p$-values for $H_0: \mu_{\text{AUC}} \leq 0.5$ and $H_1: \mu_{\text{AUC}} > 0.5$ are given for each method. 
			\label{fig:methodComparison}}
		\end{figure}

	We finally compare our approach to iRafNet~\cite[]{PMID26072483} and PLSNET~\cite[]{PMID28031031}. In Figure~\ref{fig:methodComparison}, the AUCs for PLSNET are computed using the method's default parameters, and the AUCs for iRafNet and the posterior PLS-VIP-based approach are computed using uninformative priors. Even with uninformative prior parameters, our approach still produces AUCs that are slightly but significantly above 0.5, while the AUCs of the other methods tend to be smaller. By using an input prior network and simulations based on those networks, our approach is able to better recover the true interactions of the network.

\section{Discussion}\label{sec:conclusion}

We have presented a semi-supervised approach for GRN reconstruction that can be used to refine partially known GRNs based on time-course gene expression data. In particular, we applied the PLS-VIP method to time course data by assuming that the expression or change in expression of a gene at a time point is dependent on the expression of its regulators at the previous time point. To evaluate whether each VIP score is evidence of a network edge, we developed a simulation framework that incorporates previously known regulatory relationships to model the expected gene expression dynamics, thus establishing reference distributions to which the data-derived VIP scores can be compared.  This approach directly relates the network structure to the gene expression dynamics, and the semi-supervised approach of using a prior network enables the method to recover known edges while also discovering novel edges and detecting anomalous connections. The posterior edge probabilities that are estimated for each pair of genes can be used to guide and prioritize further experiments to validate the suggested connections.

To be useful for further biological studies, GRN reconstruction methods must be able to accurately identify novel regulatory interactions that can be experimentally verified. We have shown that our semi-supervised approach is able to recover extensive portions of the regulatory dependencies of an input network, as evidenced by the high AUCs corresponding to edge recovery for certain ranges of prior probability values when applied to the DREAM and insufficient sleep datasets. By recovering known edges at a rate better than chance, we can identify novel relationships with higher confidence. More specifically, prior non-edges with high posterior edge probabilities and prior edges with low posterior edge probabilities can be treated as putative evidence for novel and anomalous edges in the network, respectively. By incorporating the putative structure of the rest of the network, the method takes into account the local regulatory relationships between genes as well as the global features of the network and regulatory dynamics when deriving these posterior probabilities. We also showed that our method was capable of novel edge detection by removing true edges from the DREAM networks and attempting to recover those edges. Similarly, by adding false edges and using the method to identify them, we showed that our method can potentially be used for anomaly detection. We also showed that our approach can outperform other related methods.

Additional extensions can be made to our approach to better model the underlying gene regulatory dynamics and potentially improve link prediction performance. For example, the modifications to PLS-VIP for time course data assumed that the expression or change in expression were linear functions of the expression at the previous time point. While this was a straightforward extension that led to good performance, other temporal modifications can be incorporated to take in account more realistic regulatory dynamics. Since genes are known to regulate the expression of other genes by its products and the generation of those products and related processes can require different amounts of time, other time course-based methods have included the expression at multiple previous time points as predictors. Other methods have included specific time points by identifying an optimal delay in response for different pairs of genes. Also, when modeling with PLSR, we assumed that if an edge exists between a pair of genes, then the connection between them is always active, which may not be the case for true regulatory dynamics. In addition, when simulating expression data and sampling edge weights for these connections, we assumed that all of the weights were concentrated around one value instead of having separate parameters for different pairs of genes. Lastly, we assumed that the response of a gene to its predictors is locally linear, so that it can be modeled (to a first approximation) by (\ref{eq:exprModel}) or (\ref{eq:deltaExprModel}).  Other network reconstruction methods have used non-linear functions to model regulatory dynamics, and a similar modification here may make our approach more representative of the underlying biology.

Yet even in spite of the simplifying assumptions, we note that our method was able to recover many of the true edges of the input prior networks as well as identify novel and anomalous edges when they were introduced into the DREAM networks. These results suggest its ability to derive posterior networks that accurately reflect gene expression dynamics and can be used to guide and prioritize further experiments and analyses.


\section*{Acknowledgements}

This work was supported by the James S.\, McDonnell Foundation (grant \#220020394) and the National Science Foundation (DMS-1547394).


\bibliographystyle{unsrt}
\bibliography{bibliography}


\clearpage

\setcounter{figure}{0}
\setcounter{equation}{0}
\setcounter{section}{0}

\renewcommand\thefigure{A-\arabic{figure}}
\renewcommand\theequation{A-\arabic{equation}}

\begin{center}
	\textbf{\large Supplementary information}
\end{center}

\section{Method details}

For each possible edge $e_{ji}$ in a network, we compute the edge probability
	\begin{equation*}
		P(e_{} = 1 \vert v_{ji}^{\text{expr}}, G^{\text{prior}})\,,
	\end{equation*}
where $v_{ji}^{\text{expr}}$ is the data-derived VIP score for the potential regulator gene $j$ in the expression model for gene $i$ and $G^{\text{prior}}$ is the prior network consisting known interactions. Applying Bayes' theorem, we have
	\begin{equation} \label{eq:bayesApp-s}
	 	P\left(e_{ji} = 1 \vert v_{ji}^{\text{expr}}, G^{\text{prior}}\right) = \frac{P\left(v_{ji}^{\text{expr}} \vert e_{ji}=1, G^{\text{prior}}\right) P\left(e_{ji}=1 \vert G^{\text{prior}}\right)}{\sum_{k \in \lbrace 0,1\rbrace} P\left(v_{ji}^{\text{expr}} \vert e_{ji}=k, G^{\text{prior}}\right) P\left(e_{ji}=k \vert G^{\text{prior}}\right)}\,,
	\end{equation}
and
	\begin{multline} \label{eq:mcIntegral-s}
		P\left(v_{ji}^{\text{expr}} \vert e_{ji}=k, G^{\text{prior}}\right) = \int_{\mathcal{X}} \int_{\mathcal{W}}P\left(v_{ji}^{\text{expr}} \vert V_{ji}\left(x^0, W\right)\right)
\\ \cdot
p(W \vert e_{ji}=k, G^{\text{prior}})\, p(x^0)\, dW\, dx^0 \;,
	\end{multline}
where $\mathcal{W}$ consists of the possible edge weights, $\mathcal{X}$ consists of the possible initial expression values, and $V_{ji}\left(x^0, W\right)$ are null VIP scores computed from expression data based on an initial condition $x^0$ and edge weights $W$. The integral (\ref{eq:mcIntegral-s}) can be estimated with Monte Carlo integration by sampling initial conditions according to $p(x^0)$ and edge weights according to $p\left(W \vert e_{ji}=k, G^{\text{prior}}\right)$ and then computing a null VIP score $V_{ji}\left(x^0, W\right)$.  We detail the key components of this procedure here.

	\subsection{Prior edge probabilities}

In equation (\ref{eq:bayesApp-s}), the prior edge probabilities $P\left(e_{ji}=1 \vert G^{\text{prior}}\right)$ must be specified for every pair of genes $i,j$ and for $k \in \lbrace0,1\rbrace$.  To reduce the number of parameters, we assume that the prior probability of an edge only depends on its existence in the observed network so that
	\begin{equation}
		P\left(e_{ji}=k \vert G^{\text{prior}}\right) = P\left(e_{ji}=k \vert e_{ji}^{\text{prior}}=l\right)
	\end{equation}
for $l \in \lbrace0,1\rbrace$. We also assume that for fixed values of $k$ and $l$, the probabilities are the same for all $i$ and $j$, leaving two parameters $p,q$ to specify:
	\begin{align}
		p &= P\left(e_{ji}=1 \vert e_{ji}^{\text{prior}}=1\right) \nonumber\\
		q &= P\left(e_{ji}=1 \vert e_{ji}^{\text{prior}}=0\right) .
	\end{align}
$p$ is therefore the probability that there is a true edge between an ordered pair of genes given that an edge exists in the reference network, and $q$ is the probability that there is a true edge given that there is not an edge in the reference network.

Since pathway databases such as KEGG contain regulatory interactions that have been expertly curated from scientific studies, the prior probabilities should reflect a high level of confidence for the edges that appear in the database-derived graph $G^{\text{prior}}$. Heuristically, $p$ should then be assigned a high value. The value to be used for $q$ depends on the interpretation of the absence of a non-edge in a pathway. If a non-edge represents no relationship between a pair of genes, then $q$ should be set to a very low value. On the other hand, if a non-edge corresponds to an unknown relationship, then a value of $q$ approaching 0.5 (reflecting even prior odds that a true relationship exists) should be used to facilitate the discovery of potential links that have yet to be studied.

	\subsection{Simulated data}

To simulate data for generating the null distributions of VIP scores, we can extend the time course models to account for gene expression stochasticity. Expression data is generated using
	\begin{align} \label{eq:exprStochastic}
		x(t+\Delta t) &= Wx(t) + \epsilon,
	\end{align}
or
	\begin{align} \label{eq:deltaExprStochastic}
		x(t+\Delta t) &= x(t) + Wx(t) + \epsilon,
	\end{align}
depending on the time course model being fit (cf. Equations (1) and (2) in the main manuscript), where $\epsilon \sim N(0, \sigma_\epsilon^2)$. While these are simple models of gene expression, they satisfy the linearity assumptions of PLSR, and thus the true covariates of those models should be identified in a predictable manner by PLS-VIP. In particular, by using a linear model to generate data, we expect PLS-VIP to compute null distributions of VIP scores that are concentrated around high and low values for the edges and non-edges of these models, respectively. Other models of gene expression may be used, such as Hill kinetics. However, the VIP scores that are computed from the generated data for the edges and non-edges of these models may not necessarily separate in the same manner as that of a linear model.

Probability distributions also need to be specified for $p\left(x^0\right)$ and $p\left(W \vert e_{ji}=k, G^{\text{prior}}\right)$. Since microarray gene expression data is typically assumed to be normally distributed (following the customary $\log_2$ transformation), the initial expression vector $x(0)=x^0$ is drawn according to $x^0 \sim N\left(0, \sigma_x^2\right)$. To assign weights in $W$, we first assume that in the conditional of $p\left(W \vert e_{ji}=k, G^{\text{prior}}\right)$, knowledge of $e_{ji}$ supersedes the observation $e_{ji}^{\text{prior}}$. Zero weights are then assigned to the non-edges in the graph. For the edges in the graph, weights are drawn from an even mixture of two normal distributions, $N\left(\mu_w, \sigma_w^2\right)$ and $N\left(-\mu_w, \sigma_w^2\right)$, so that the graph contains a mixture of up- and down-regulatory edges.

	\subsection{Integral estimation}

Combining the previous assumptions, we can now estimate the integral (\ref{eq:mcIntegral-s}) with Monte Carlo integration. We first compute the data-derived VIP scores $v_{ji}^{\text{expr}}$ from an input expression data set for all ordered pairs of genes. Next, for each ordered pair of genes $\left(j,i\right)$, $k \in \lbrace 0,1\rbrace$, and for a large number of iterations $N$, we sample weights $W$ according to $p\left(W \vert e_{ji}=k, G^{\text{prior}}\right)$ and initial conditions $x^0$ according to $p\left(x^0 \right)$. Data is then simulated from the weights and initial conditions using one of the time course models (\ref{eq:exprStochastic}) or (\ref{eq:deltaExprStochastic}), from which VIP scores are then computed for $\left(j,i\right)$. The scores are then aggregated over the $N$ iterations to form a null distribution for each $i,j,k$, and (\ref{eq:mcIntegral-s}) is estimated by
	\begin{multline} \label{eq:integralEstimation}
		\int_{\mathcal{X}} \int_{\mathcal{W}}P\left( v_{ji}^{\text{expr}} \vert V_{ji}\left(x^0, W\right)\right) p(W \vert e_{ji}=k, G^{\text{prior}}) p(x^0) dW dx^0  \\ \approx  \begin{cases}
			r & \text{if } k=1\\
			1-r & \text{if } k=0,
		\end{cases}
	\end{multline}
where $r$ is the fraction of null VIP scores that are less than or equal to the data-derived VIP score $v_{ji}^{\text{expr}}$. The estimates of (\ref{eq:mcIntegral-s}) can then be plugged into \ref{eq:bayesApp-s} to obtain the posterior probability of an edge from gene $i$ to gene $j$. In (\ref{eq:integralEstimation}), the two cases have been specified so that the posterior probability (\ref{eq:bayesApp-s}) retains the property that higher VIP scores can still contribute to a higher edge probability.

	\subsection{Computational considerations}

One of the major drawbacks of our method is its computational complexity. Currently, the method computes posterior edge probabilities for every pair of genes, and for each of those pairs, two distribution of VIP scores corresponding to the edge and non-edge cases of that pair need to be generated. For a reliable estimate of the probability, a large number of iterations should be used. For each of these iterations, PLSR and VIP computations need to be performed. With the NIPALS algorithm for PLSR, the number of operations will depend on the the number of available samples, and the number of predictor genes, and the number of PLS components that are used. Due to the large number of operations, without any restrictions or modifications, the method does not lend itself to applications on large networks. To improve the overall computational performance, other PLSR algorithms such as SIMPLS or kernel PLS may be used in place of the NIPALS algorithm. Instead of computing posterior edge probabilities for every ordered pair of genes, one may also identifying components, clusters, or other subnetworks as a preprocessing step and then applying the method to these subnetworks. This avoids having to compute probabilities for genes that are distant in the network and unlikely to interact.

\section{Validation framework}

To evaluate the method's ability to detect novel edges, we randomly remove edges (blue) from the DREAM networks, apply the method to estimate the posterior edge probabilities of the non-edges of the modified graphs, and calculate an AUC by ranking these posterior edge probabilities. This AUC can be interpreted as the probability that a randomly chosen missing edge has a higher posterior edge probability than that of randomly chosen a non-edge. Similarly, to evaluate the method's ability to detect anomalous edges, we randomly add edges (red) to the DREAM networks, apply the method to estimate the posterior edge probabilities of the edges of the modified graphs, and calculate an AUC by ranking thesea posterior edge probabilities. In this case, the AUC can be interpreted as the probability that a randomly chosen edge has a higher posterior edge probability than that of a randomly chosen false edge.

\begin{figure}[hbt]
	\centering
	\includegraphics[width=\textwidth]{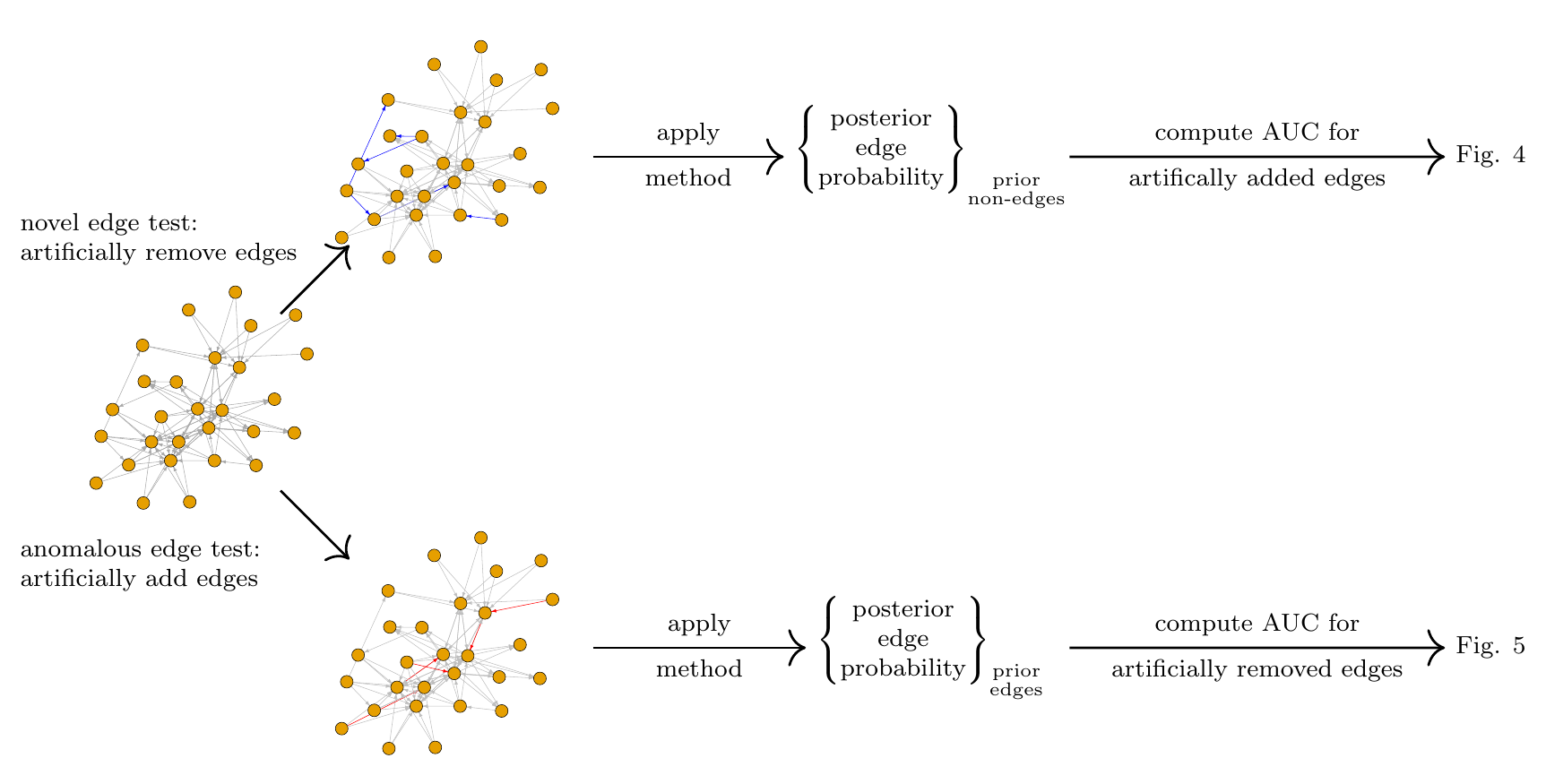}
	\caption{Novel and anomalous edge detection evaluation.}
	\label{fig:novelAnomalousWorkflow}
\end{figure}

\section{Choice of parameter $\mu$ for gene dynamics simulation}

We consider the effect of the parameter $\mu$ that is used in simulating the synthetic time course profiles. In these simulations, $\mu$ is the average strength of a regulatory connection between a pair of genes. For small values of $\mu$, regulators will weakly affect their targets, so the effect of a regulator may be indistinguishable from noise in the null models. For each ordered pair of genes, this should manifest in an edge VIP distribution that highly overlaps with the corresponding non-edge VIP distribution. As $\mu$ increases, the regulatory strength increases, so PLS-VIP should be able to better identify a gene's regulators in the simulations and assign higher null VIP scores to those regulators. The edge VIP distributions should then shift to larger VIP values and show less overlap with the corresponding non-edge VIP distributions. However, if $\mu$ is too large, the data-derived VIP scores may be small relative to the null edge VIP scores, causing prior edges to be assigned small posterior edge probabilties.

\begin{figure}
	\centering
	\subfigure[\label{sf:aucVp-CR}]{\includegraphics[width=.496\textwidth]{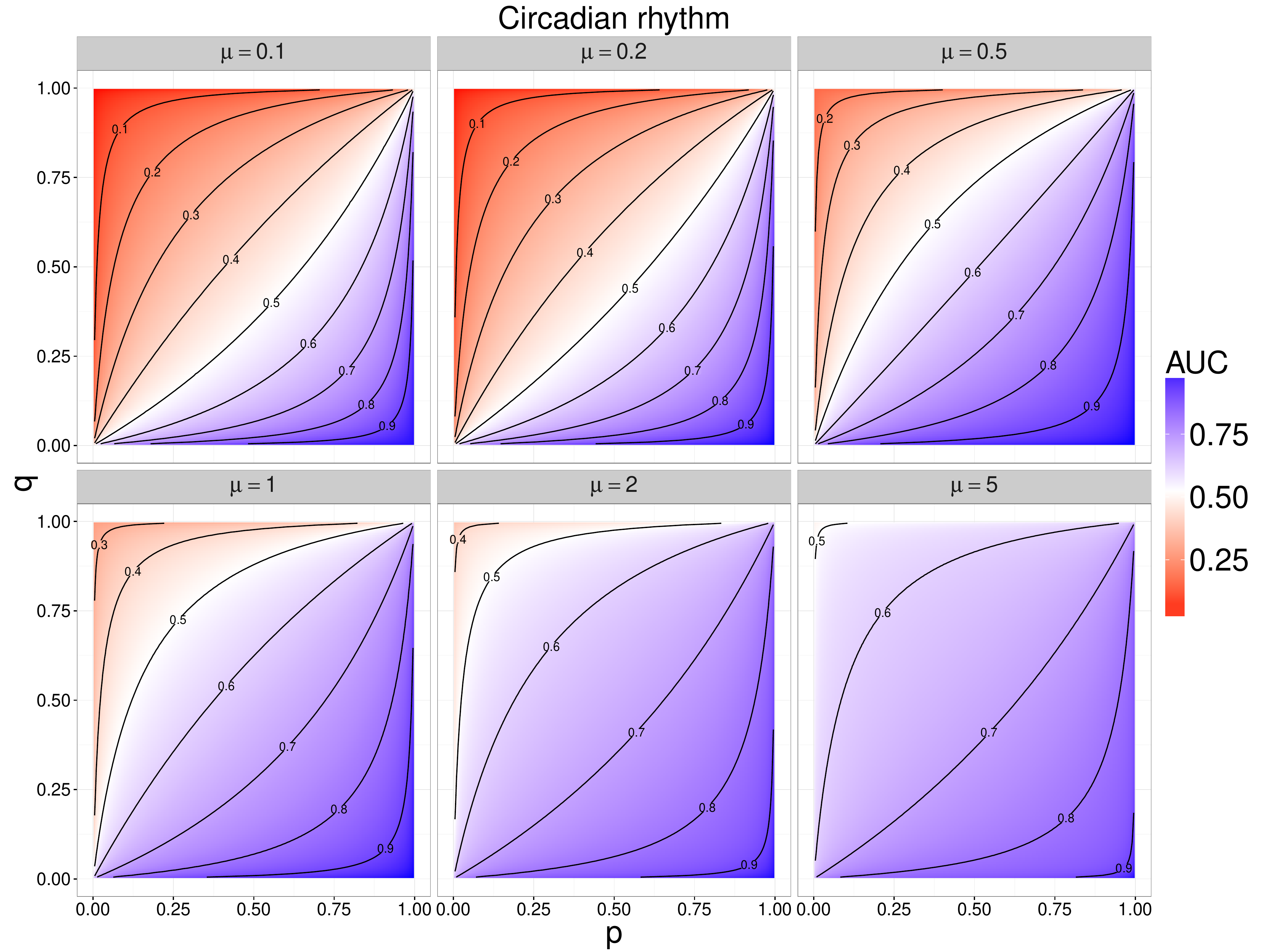}}
	\subfigure[\label{sf:aucVp-FAM}]{\includegraphics[width=.496\textwidth]{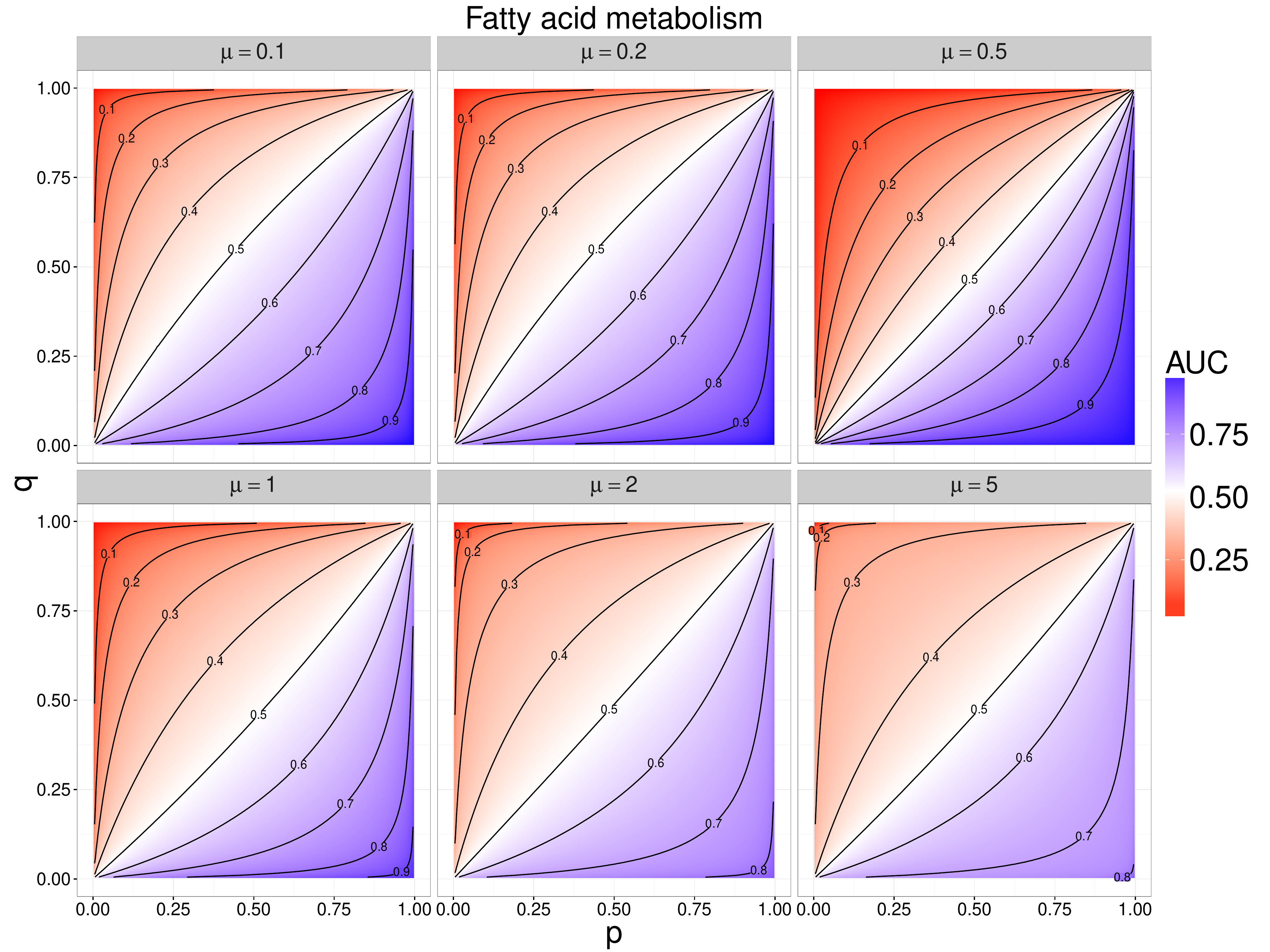}}
	\caption{AUCs at different values of $\mu$, $p$, and $q$ for the \subref{sf:aucVp-CR} circadian rhythm and \subref{sf:aucVp-FAM} fatty acid metabolism pathways.
	The range of observable AUCs and the rate of change in the AUC with respect to $\mu$ at fixed values of the $p$ and $q$ can vary between different pathways.\label{fig:aucVp}}
\end{figure}

\begin{figure}
	\centering
	\subfigure[\label{sf:roc-CR}]{\includegraphics[width=.496\textwidth]{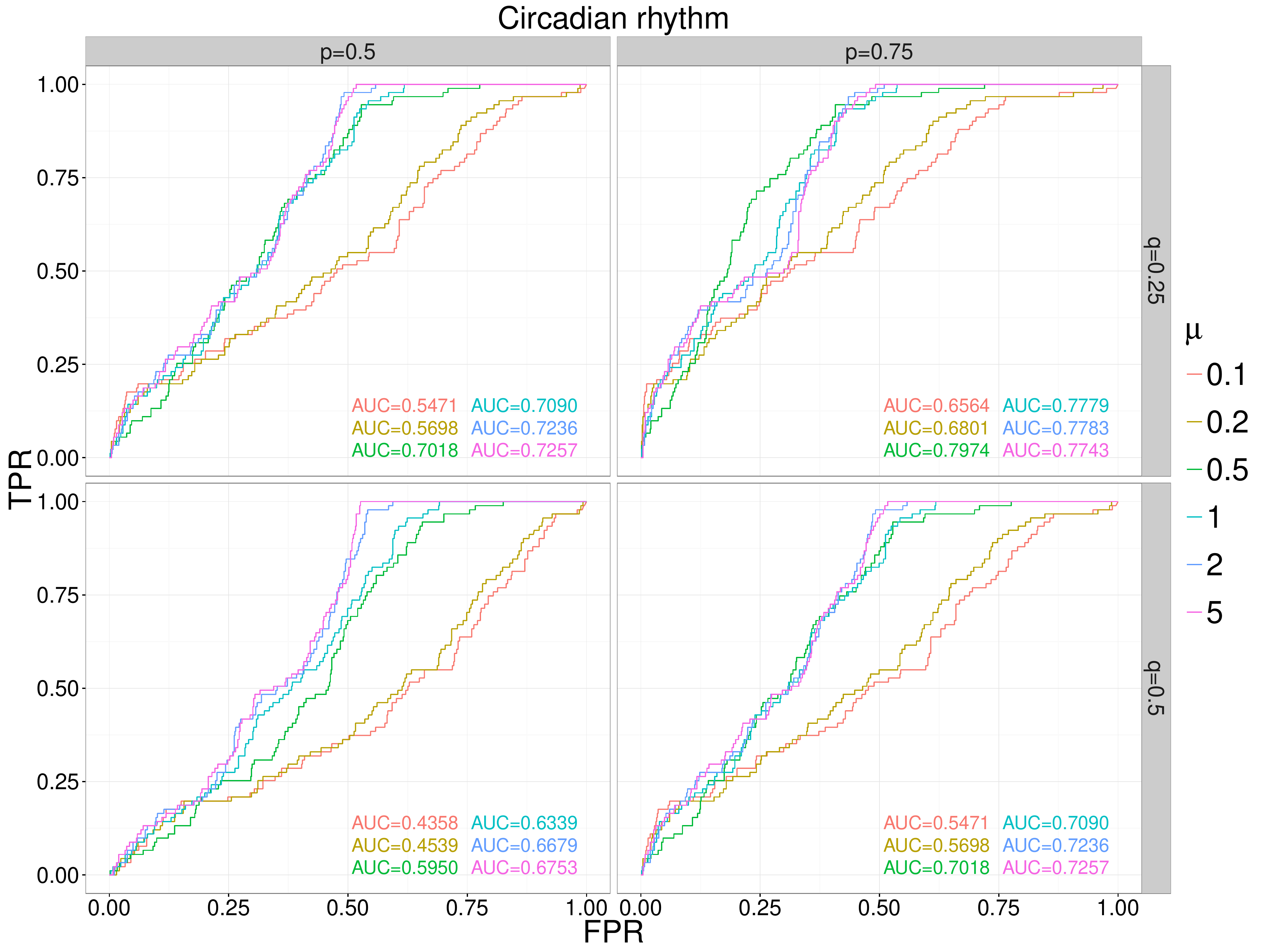}}
	\subfigure[\label{sf:roc-FAM}]{\includegraphics[width=.496\textwidth]{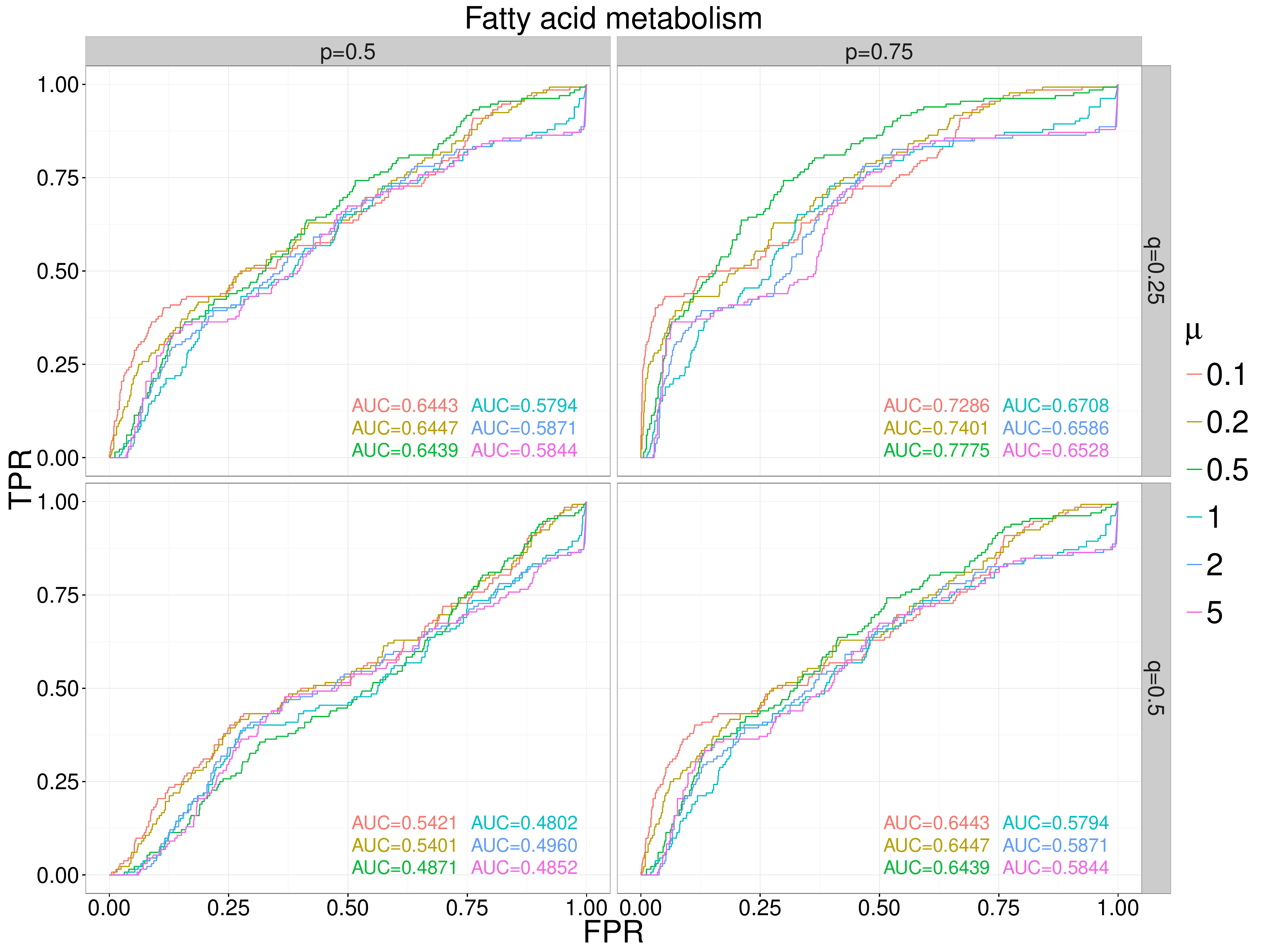}}
	\caption{ROC curves at different values of $\mu$, $p$, and $q$ for the \subref{sf:roc-CR} circadian rhythm and \subref{sf:roc-FAM} fatty acid metabolism pathways.
	\label{fig:rocCurves}}
\end{figure}

\begin{figure}
	\centering
	\includegraphics[width=\textwidth]{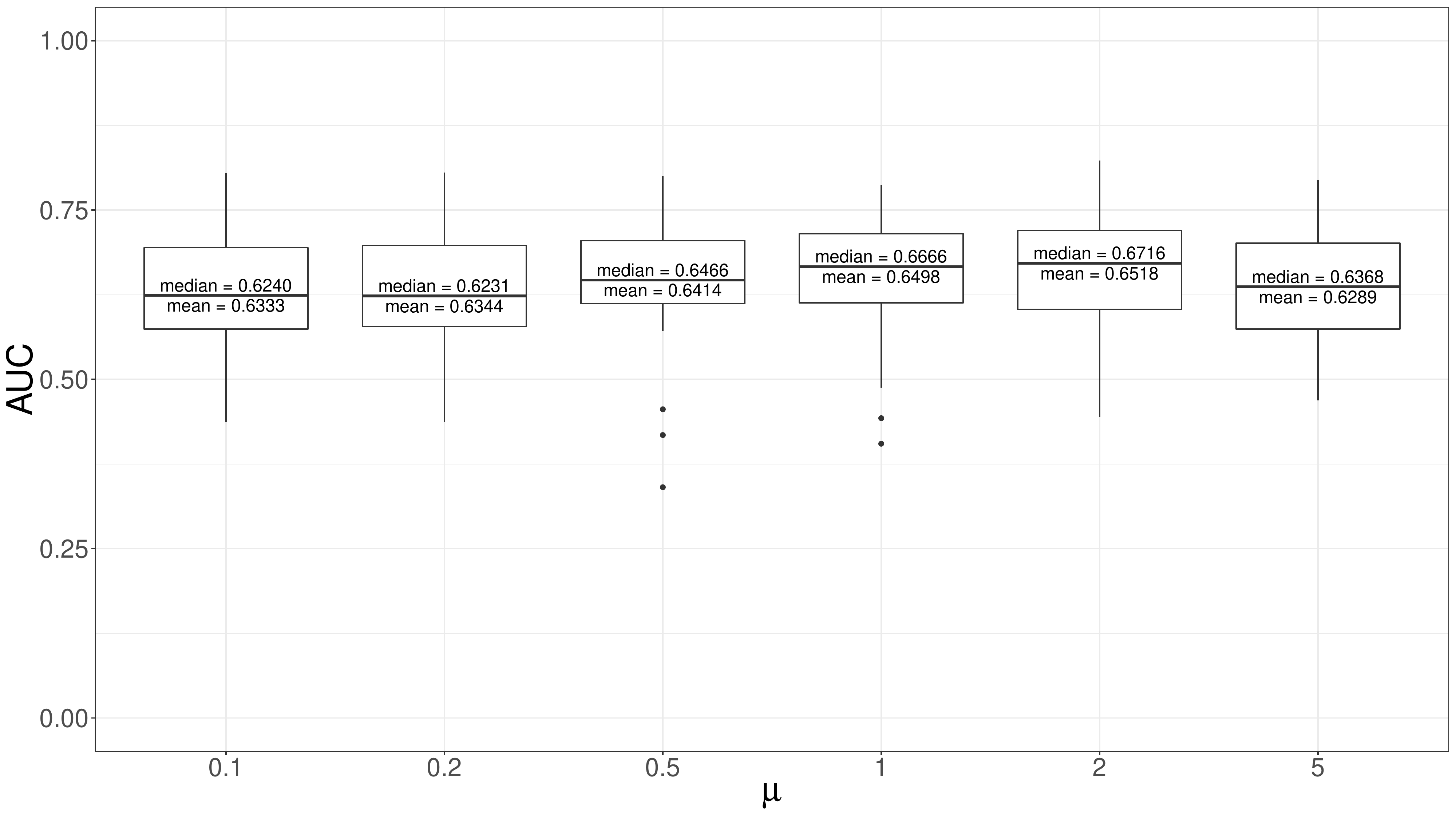}
	\caption{AUCs at different values of $\mu$ with $p=q=0.5$ for the PLS-VIP-based posterior edge probabilities when applied to the DREAM datasets.
	\label{fig:dreamAUCvMu}}
\end{figure}

In Figure~\ref{fig:aucVp}, heatmaps of the AUCs as $p$ and $q$ are varied at fixed values of $\mu$ are shown for the circadian rhythm and fatty acid metabolism pathways, and contours are placed at increments of 0.1. In Figure~\ref{fig:rocCurves}, ROC curves are shown for selected values of the three parameters. For our simulations, we set $\sigma_x = \sigma_w = 0.1$ so that at $\mu=1$, there is little to no overlap between the distributions of up- and down-regulatory weights. For both pathways, we again see that the AUC increases with increasing $p$ and decreasing $q$. However, the behavior differs between the two pathways with changes in $\mu$. For the circadian rhythm pathway, the AUC initially increases with $\mu$ at almost all values of the prior parameters. In particular, based on the gap in the ROC curves at the selected values of $p$ and $q$, a large change in AUC occurs between $\mu = 0.2$ and $\mu = 0.5$. As $\mu$ increases to 5, the AUC decreases for some values of $p$ and $q$, indicated by the shift in some of the heatmap contours towards the lower right corner. In addition, the spread between the contours increases, so the AUC landscape flattens and the AUC becomes more robust to changes to $p$ and $q$ when neither parameter is close to 0 or 1. In contrast, the ROC curves highly overlap for the fatty acid metabolism pathway, and while the AUC heatmaps and contours change with increasing $\mu$, after $\mu=1$, the changes are very gradual, so the AUC appears to be robust to changes in $\mu$. In addition, the range of AUC values that are observed for large $\mu$ are different that those seen with the circadian rhythm pathway. Given the difference in behavior in the AUC with respect to $\mu$ between the two pathways, a recommended choice of $\mu$, such as one that maximizes the AUC at fixed values of the prior parameters in order to recover most of the partially known network, can depend on the pathway, the expression time course profiles and derived VIP scores, and the choice of prior parameters. In Figure~\ref{fig:dreamAUCvMu}, boxplots of the AUCs for the DREAM datasets are shown for $\mu \in \lbrace 0.1,\,0.2,\,0.5,\,1,\,2,\,5\rbrace$. Since the performance on the synthetic datasets appears to be best around $\mu=1$ and $\mu=2$, we chose $\mu=1$ for the results of the sleep dataset.

\section{Automorphically equivalent nodes}

\begin{figure}
	\centering
	\subfigure[\label{sf:pVv-CR}]{\includegraphics[width=.496\textwidth]{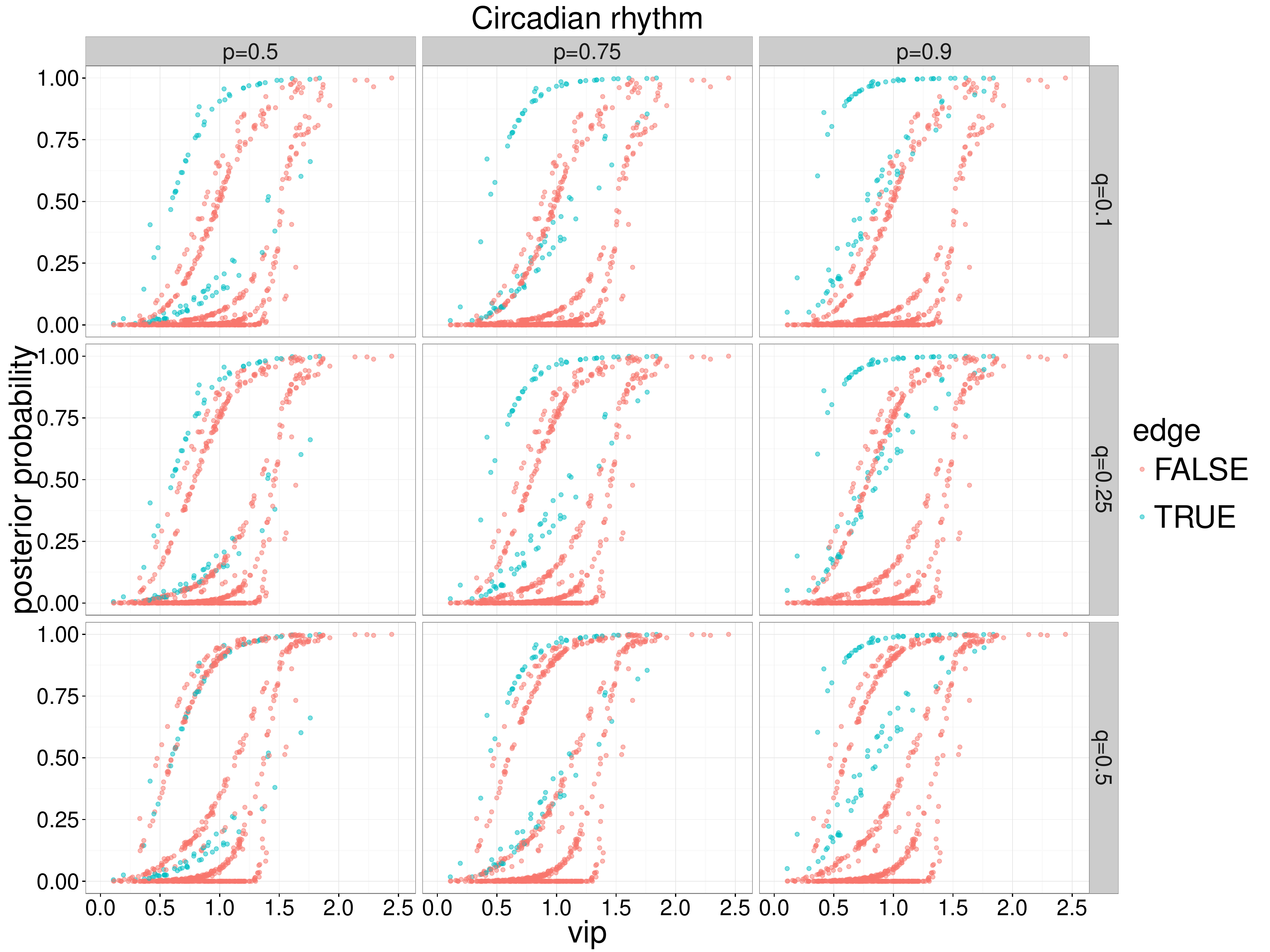}}
	\subfigure[\label{sf:pVv-FAM}]{\includegraphics[width=.496\textwidth]{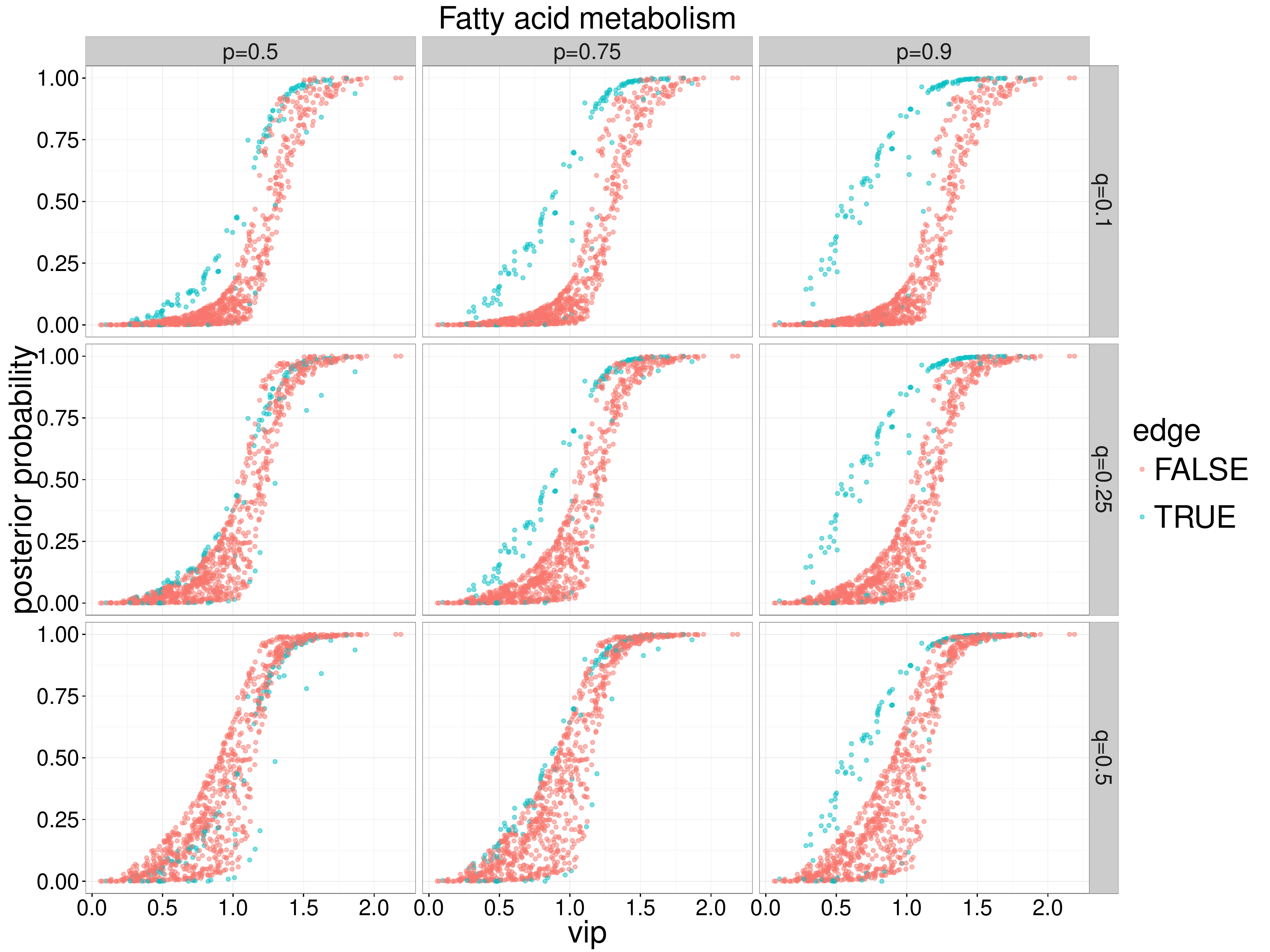}}
	\caption{Posterior edge probability vs. VIP scores for each pair of genes at fixed values of $p$ (columns) and $q$ (rows) for the (a) circadian rhythm and (b) fatty acid metabolism pathways. Posterior probabilities increase with $p$ and $q$, and large VIP scores tend to have higher posterior probabilities, but a small VIP can still result in a high posterior probability and vice-versa. The structure of the input pathway can also influence the posterior probabilities.\label{fig:postVvip}}

	\subfigure[\label{sf:network-CR}]{\includegraphics[width=.496\textwidth]{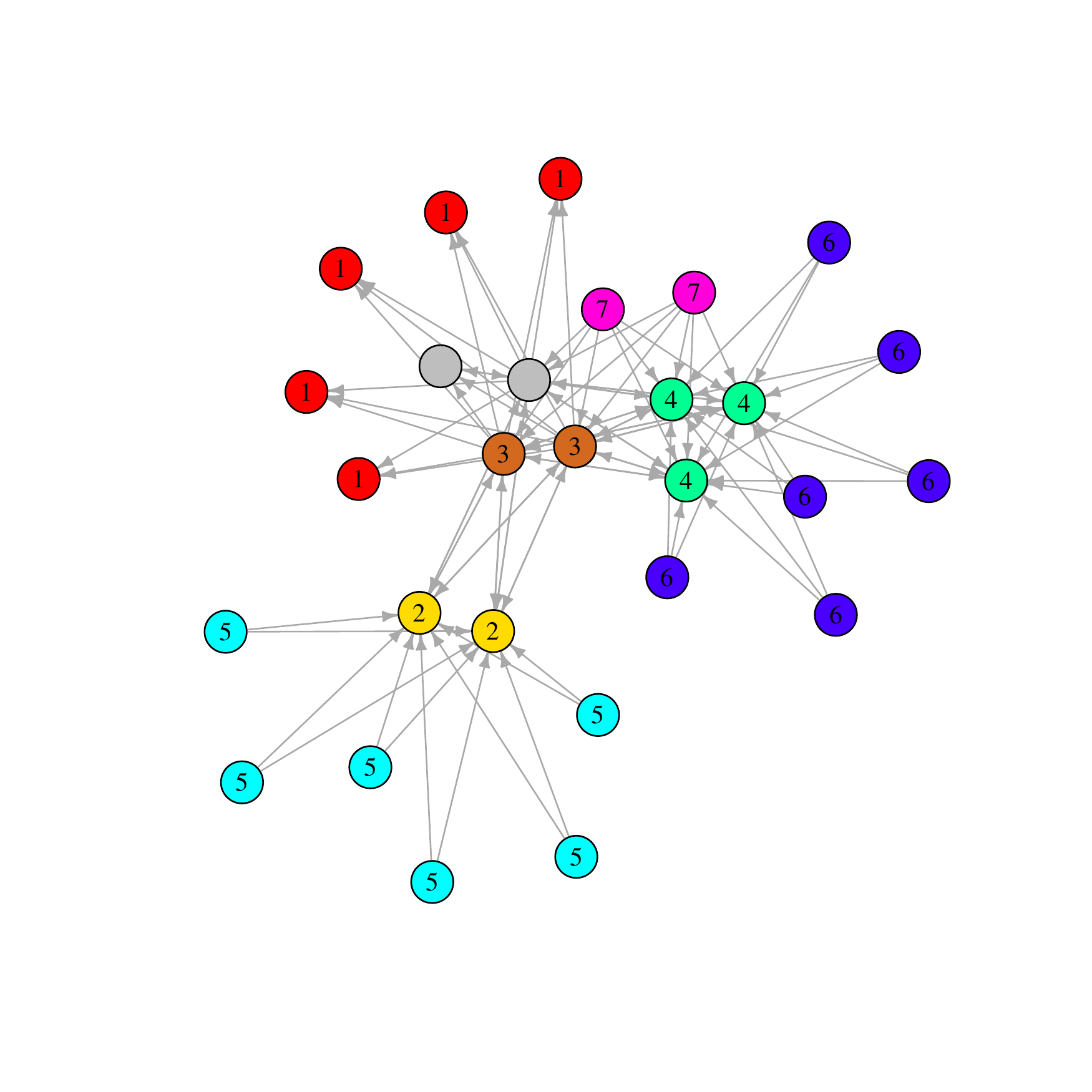}}
	\subfigure[\label{sf:network-FAM}]{\includegraphics[width=.496\textwidth]{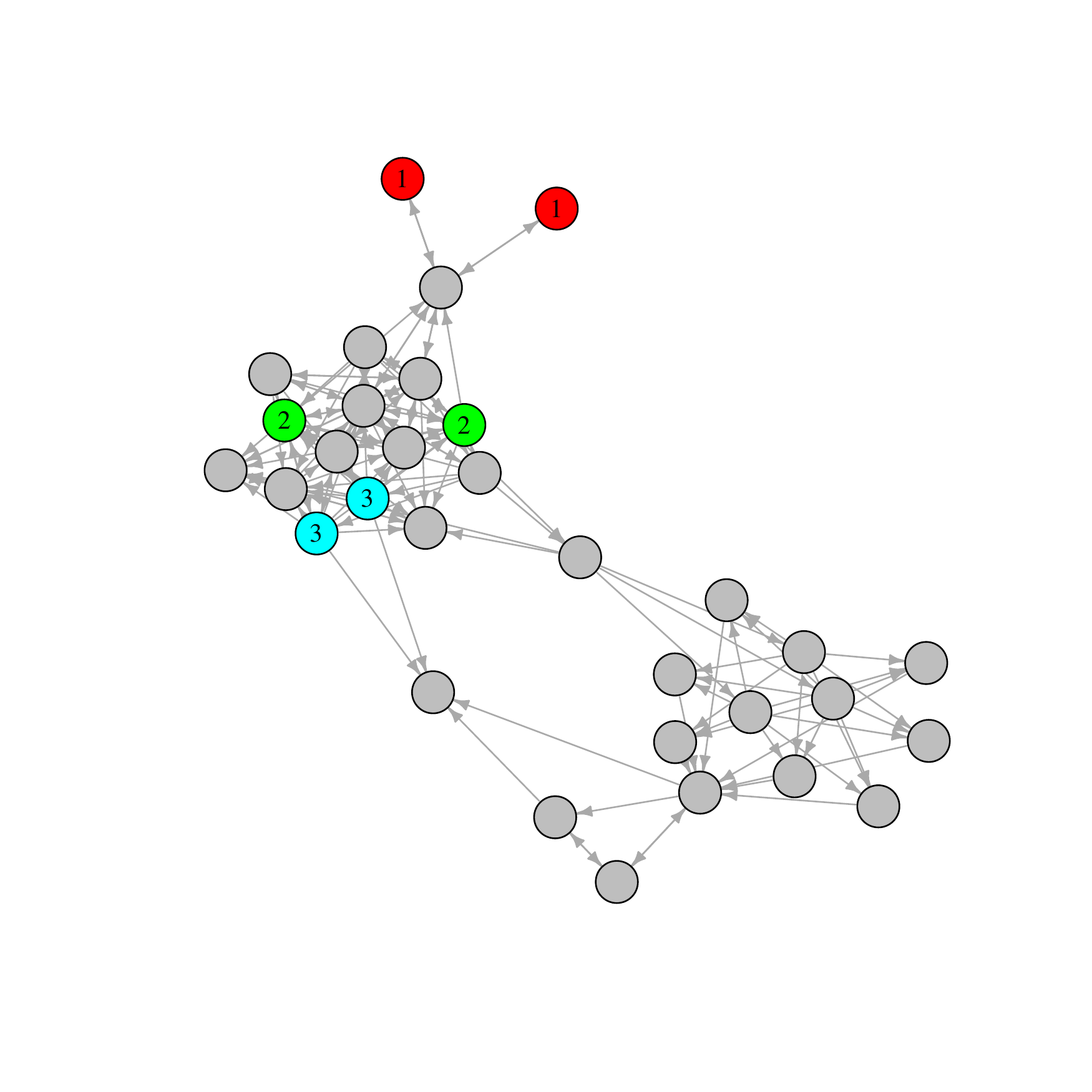}}
	\caption{Network plots of the \subref{sf:network-CR} circadian rhythm and \subref{sf:network-FAM} fatty acid metabolism pathways. \label{fig:networks}}
\end{figure}

In Figure~\ref{fig:postVvip}, the posterior probabilities for each ordered pair of genes are plotted against the VIP scores for the circadian rhythm and fatty acid metabolism pathways at a variety of prior probability values $p$ and $q$. It can be observed that the relationship between the VIP score and the posterior probability falls into one of several bands, with the bands being more pronounced in the circadian rhythm network. We can attribute this pattern to the existence of automorphically equivalent nodes in the network.

An automorphism $f$ of a graph is a mapping of the graph back to itself that preserves edges, i.e., $(u,\,v)$ is an edge if and only if $f(u),\,f(v)$ is an edge. Two nodes $u$ and $v$ are said to be automorphically equivalent if there exists an automorphism on the graph that maps $u$ to $v$.

Considering only the structure of the network, nodes that are in the same equivalence class essentially play the same role in the network and affect nodes in other classes in the same manner. Under the assumptions used for the time course simulations, when the method samples different initial conditions and edge weights over many Monte Carlo iterations, the method will then produce distributions of edge (non-edge) VIP scores that are similar for edges (non-edges) between the same equivalence classes; in the limit of a large number of iterations, the distributions will be the same. A pair of VIP score distributions that is representative of the edges or non-edges between two equivalence classes then defines a curve in the posterior probability vs. VIP score plot. This suggests that the method is able to take into account not only local regulatory relationships between genes but also global network features and gene expression dynamics in order to derive posterior probabilities.

In Figure~\ref{fig:networks}, plots of the two networks are shown. For each pathway, nodes belonging to singleton node equivalence classes are grey and unlabeled; otherwise, nodes are colored and labeled by the class that they belong to. Since the fatty acid metabolism pathway has many more equivalence classes that are also singletons, the banding behavior that appears with the circadian rhythm pathway is much less apparent in the fatty acid metabolism pathway.

\end{document}